\documentclass[english]{elsarticle}
\usepackage[T1]{fontenc}
\usepackage[latin9]{inputenc}
\usepackage{geometry}
\usepackage{amsmath}
\usepackage{amssymb}
\usepackage{graphicx}
\usepackage{esint}
\makeatletter
\journal{Chemical Physics}

\makeatother

\usepackage{babel}
\begin{document}

\title{Excitation Dynamics and Relaxation in a Molecular Heterodimer}

\author[vu,fi]{V.\ Balevi\v{c}ius,\ Jr.}

\author[vu,fi]{A.\ Gelzinis}

\author[vu,chi]{D.\ Abramavicius}

\author[cuni]{T.\ Man\v{c}al}

\author[vu,fi]{L.\ Valkunas\corref{cor}}

\ead{leonas.valkunas@ff.vu.lt}

\address[vu]{Department of Theoretical Physics, Faculty of Physics, Vilnius University,
Sauletekio Avenue 9, build. 3, LT-10222 Vilnius, Lithuania}

\address[fi]{Center for Physical Sciences and Technology, Institute of Physics,
Savanoriu Avenue 231, LT-02300 Vilnius, Lithuania}

\address[chi]{State Key Laboratory of Supramolecular Structure and Materials, Jilin
University, 2699 Qianjin Street, Changchun 130012, PR China}

\address[cuni]{Faculty of Mathematics and Physics, Charles University in Prague,
Ke Karlovu 5, CZ-121 16 Prague 2, Czech Republic}

\cortext[cor]{Corresponding author at: Department of Theoretical Physics, Faculty
of Physics, Vilnius University, Sauletekio Avenue 9, build. 3, LT-10222
Vilnius, Lithuania}
\begin{abstract}
The exciton dynamics in a molecular heterodimer is studied as a function
of differences in excitation and reorganization energies, asymmetry
in transition dipole moments and excited state lifetimes. The heterodimer
is composed of two molecules modeled as two-level systems coupled
by the resonance interaction. The system-bath coupling is taken into
account as a modulating factor of the energy gap of the molecular
excitation, while the relaxation to the ground state is treated phenomenologically.
Comparison of the description of the excitation dynamics modeled using
either the Redfield equations (secular and full forms) or the Hierarchical
quantum master equation (HQME) is demonstrated and discussed. Possible
role of the dimer as an excitation quenching center in photosynthesis
self-regulation is discussed. It is concluded that the system-bath
interaction rather than the excitonic effect determines the excitation
quenching ability of such a dimer.\end{abstract}
\maketitle

\section{Introduction}

Photosynthetic pigment-protein complexes arranged as light-harvesting
antenna absorb the Sun light and transfer the captured energy towards
the photosynthetic reaction center, where electron transfer across
the membrane is initiated. In addition light harvesting antennae also
carry out self-regulation function, which is a physiologically highly
significant strategy evolved by plants \citep{A.V.Ruban2011}. Thanks
to the excitation energy density control in photosystem II, termed
non-photochemical quenching (NPQ), plant photosynthesis can function
efficiently under very different light conditions, from low up to
very high intensities corresponding to daily changes of illumination
at the same location. The NPQ phenomenon is usually attributed to
some activated quenching species which allow the excitation to undergo
a rapid non-radiative decay. The exact location of the quencher within
the antenna and its precise nature are a matter of on-going debate,
with both chlorophylls (Chl) and carotenoids (Car) being put forward
as essential components of the quenching mechanism. The promising
candidates for the quenching mechanisms (in no order of preference)
are: (i) quenching by a Chl\textendash{}Chl dimer showing the charge
transfer (CT) state character \citep{M.G.Mueller2010} or via the
formation of Chl\textendash{}Chl excimeric states \citep{P.Horton1996};
(ii) by a CT state between Chl and Car (xanthophyll) resulting in
generation of the cation radical state of the xanthophyll \citep{holt-flemingScience05,T.K.Ahn2008};
(iii) via the excitonic coupling of Chl to a short-lived xanthophyll
excited state \citep{S.Bode2009}, and (iv) by the direct energy transfer
from the Chl pool to a particular Car (lutein) \citep{vanGrondelle-NATURE2007}.
Both latter suggestions are based on the fact that the lifetime of
the lowest excited state of the lutein molecule is short (<10 ps).
Evidently, this could explain the possible quenching efficiency in
the light-harvesting complex of photosystem II (LHCII) if a specific
arrangement between the Chl and Car molecules is established under
the NPQ conditions. Thus, according to the latter suggestions, the
simplest candidate which might be responsible for the NPQ, is an asymmetric
molecular dimer (a heterodimer). The possibility to attribute the
excitation quenching ability to the Chl\textendash{}Car heterodimer
was also suggested and experimentally studied for model dyads composed
of Car and tetrappyroles \citep{berera-kennisPNAS06,Liao2011}. A
dimer is also the simplest molecular aggregate where the excitonic
features are well expressed.

Spectral characteristics of molecular dimers are often remarkably
different from those of the individual molecules. Their absorption
bands can be significantly shifted in comparison with those of their
constituent molecules, mainly due to intermolecular interactions resulting
in a nonlocal character of their excited eigenstates. These delocalized
eigenstates of electronic excitations are usually termed molecular
excitons \citep{Davydov-book,RashbaExcitons,Silinsh1994,valkunasbook}.
The interaction of electronic excitations with intra- and inter-molecular
vibrations causes a disruption of the phase relationship between excited
states of the molecules constituting the exciton wave functions \citep{valkunasbook,May2004}.
Such type of interaction makes a distinct influence on the exciton
dynamics, and plays the dominant role in determining the exciton relaxation
pathways.

Usually, a homodimer is used for considering various aspects of the
exciton dynamics and relaxation \citep{ChoFleming2005,Kjellberg2006,Psliakov-Fleming-JCP2006,Abramavicius2010}.
For a heterodimer, the distinctness of constituent monomers is often
limited to excitation energies \citep{G.S.Schlau-Cohen2011,Ishizaki2009}.
Some aspects which could be attributed to the heterodimer were also
disclosed by analyzing the exciton\textendash{}CT state mixing problems
\citep{Renger2004,T.Mancal2006}. Recent experiments based on coherent
photon echo measurements demonstrated the possibility to follow the
coherent phase dynamics and incoherent population relaxation of excitons
in photosynthetic pigment-protein complexes \citep{G.S.Schlau-Cohen2011,Grondelle2006,HeijsKnoester2007}.
However, possible role of the coherent dynamics in determining such
type of the regulation function \textendash{} attributing it to the
ability of the excitation quenching in heterodimers \textendash{}
has not been considered yet. Here, such type of the analysis is presented.
We consider the effects originated from the differences in the excitation
energies, reorganization energies and excitation lifetimes of the
constituent molecules of the heterodimer.

The paper is organized as follows. In section \ref{sub:Excitonic-Dimer}
we introduce a theoretical model of an excitonic dimer and discuss
the general form of equations describing the excitation dynamics in
the system. In sections \ref{sub:Redfield-Relaxation-Scheme} and
\ref{sub:HEOM-Relaxation-Scheme} we introduce two specific theories
of relaxation that are nowadays standard in the literature and in
section \ref{sub:Exciton-Decay} we discuss our model for the radiationless
excitation energy dissipation. Numerical results for the excited state
dynamics under various parameters of the dimer system with and without
relaxation to the ground state are presented in sections \ref{sub:Population-Kinetics}
and \ref{sub:Relaxation-to-the}, respectively. The consequences of
these results and the possible role of the heterodimer in determining
the NPQ are discussed in section \ref{sec:Discussion}.

\section{Modeling}

\subsection{Excitonic Dimer\label{sub:Excitonic-Dimer}}

The exciton spectra of molecular aggregates are usually considered
in the so-called Heitler-London approximation \citep{Davydov-book,RashbaExcitons,valkunasbook}.
According to this theoretical concept, monomers comprising the dimer
are characterized by two electronic levels, which reflect the resonant
transition in the spectral region under consideration. The states
of the dimer can be constructed on a Hilbert space, which is a direct
product of the Hilbert spaces of the monomers. The ground state of
the dimer is constructed from monomers $i=a$ and $b$ as $|g\rangle\equiv|g_{a}\rangle|g_{b}\rangle$.
Similarly, two singly-excited states are defined as $|i\rangle\equiv|e_{i}\rangle|g_{j}\rangle$,
$j\neq i$ where the \textit{i}th molecule is in the excited state
and the \textit{j}th molecule in the ground state. The two singly-excited
states are coupled to each other by the resonance interaction $J_{ab}$,
and consequently they are not eigenstates of the system. A double-excited
state $|f\rangle\equiv|e_{i}\rangle|e_{j}\rangle$ where both molecules
are excited is not relevant in our analysis. The given states form
the site basis.

Interaction of the excitonic system with its environment forming a
thermodynamic bath, has to be taken into account when analyzing the
exciton dynamics and relaxation. Thus, the full Hamiltonian $H$ should
contain a purely system part $H_{S}$, describing only the electronic
excitations, the bath part $H_{B}$ and the system-bath interaction
part $H_{SB}$ \citep{May2004}: 
\begin{equation}
H\equiv H_{S}+H_{SB}+H_{B}.\label{eq:total Hamiltonian}
\end{equation}
The individual terms in the site basis explicitly read:
\begin{equation}
H_{S}=\sum_{i}(\epsilon_{i}^{0}+\lambda_{i})K_{i}+J_{ab}(|a\rangle\langle b|+|b\rangle\langle a|),\label{eq:sys_ham}
\end{equation}

\begin{equation}
H_{B}=T(p)+V_{g}(q),\label{eq:bath_ham}
\end{equation}

\begin{equation}
H_{SB}=\sum_{i}\Delta V_{i}(q)K_{i}.\label{eq:sb_ham}
\end{equation}
Here, $i=a,\, b$; $\epsilon_{i}^{0}$ denotes the energies corresponding
to the electronic excitation in the \textit{i}th monomer, and $K_{i}\equiv|i\rangle\langle i|$
is the projector onto the \textit{i}th site. The system Hamiltonian
$H_{S}$ has the form of the Frenkel exciton Hamiltonian. Operators
$T(p)$ and $V_{i}(q)$ denote the kinetic energy of the nuclei and
the nuclear potential energy surface of the\textit{ i}th site, respectively;
$p$ and $q$ denote the generalized momenta and coordinates of the
bath. We defined the reorganization energy as 
\begin{equation}
\lambda_{i}=\langle V_{i}(q)-V_{g}(q)\rangle_{q},\label{eq:reorg. energy}
\end{equation}
where the angular brackets $\langle\ldots\rangle_{q}$ represent the
averaging over the equilibrium bath:

\begin{equation}
\langle\ldots\rangle_{q}=Tr_{q}\left\{ \hat{R}_{eq}\ldots\right\} .\label{eq:averaging}
\end{equation}
Here, $Tr_{q}\left\{ \ldots\right\} $ denotes the trace operation,
and $\hat{R}_{eq}$ is the canonical equilibrium density operator
of the bath:

\begin{equation}
\hat{R}_{eq}=e^{-\beta H_{B}}\big/Tr_{q}\left\{ e^{-\beta H_{B}}\right\} ,\label{eq:R_eq}
\end{equation}
where $\beta=1/k_{\textrm{B}}T$, $T$ is the temperature and $k_{\textrm{B}}$
is the Boltzmann constant. The so-called energy gap operator $\Delta V_{i}(q)$
defined as

\begin{equation}
\Delta V_{i}(q)=V_{i}(q)-V_{g}(q)-\lambda_{i}\label{eq:e_gap_func}
\end{equation}
describes the fluctuations of the energy gap between the potential
energy surfaces of the state $|i\rangle$ and the ground state $|g\rangle$.
We next denote $\epsilon_{i}\equiv\epsilon_{i}^{0}+\lambda_{i}$.

In order to find the excitonic states, which diagonalize the system
Hamiltonian, we use a unitary transformation:

\begin{equation}
|\mu\rangle=\sum_{i}(U^{-1})_{\mu i}|i\rangle,\label{eq:exc_expa}
\end{equation}
which yields energies of the exciton states: $\epsilon_{\mu}=\left(U^{-1}H_{S}U\right)_{\mu\mu}$.
In the following we use the Roman letters for the site basis and the
Greek letters for the eigenstate basis. In the case of a dimer the
transformation matrix has a convenient analytical form \citep{T.Mancal2008}:

\begin{equation}
U=\left(\begin{array}{cc}
\cos\theta & -\sin\theta\\
\sin\theta & \cos\theta
\end{array}\right),\label{eq:uni_transf}
\end{equation}
where $\theta$ is the so-called mixing angle, which is defined as

\begin{equation}
\theta\equiv\frac{1}{2}\arctan\frac{2J_{ab}}{\epsilon_{a}-\epsilon_{b}}\label{eq:mix_angle}
\end{equation}
and has values ranging from $0$ to $\pi/2$.

The most detailed description of our system of interest and its environment
is given by a density operator $\hat{W}$ acting on the joint Hilbert
space of system and bath states. However, since we are interested
only in the electronic degrees of freedom (DOF), we switch to the
reduced density operator (RDO) $\hat{\rho}$, which is obtained by
tracing the full density operator over the irrelevant (environment)
DOF as $\hat{\rho}=Tr_{q}\left\{ \hat{W}\right\} $. In the basis
of electronic states the RDO is represented by its matrix elements
$\rho_{ab}=\langle a|\hat{\rho}|b\rangle$. Usually it is assumed
that the initial full density operator $\hat{W_{0}}$ can be factorized
into the RDO, $\hat{\rho_{0}}$, and the complementary equilibrium
density operator of the bath: $\hat{W_{0}}=\hat{\rho}_{0}\otimes\hat{R}_{eq}$.
This factorization assumption is well motivated in the spectroscopic
applications by the existence of a single electronic ground state
and a large energy gap between the ground and the excited states.
Ultrafast excitation of the system by laser light corresponds to the
Franck-Condon transition (i.e. the nuclear DOF are not involved in
the process), and the state remains factorized immediately after excitation.

Using these definitions the equations of motion for the RDO are given
in a general form by \citep{May2004,T.Mancal2008}:

\begin{equation}
\frac{\partial}{\partial t}\hat{\rho}=-i{\cal L}_{S}\hat{\rho}+{\cal D}[\hat{\rho}]+{\cal K}\hat{\rho},\label{eq:general_eq}
\end{equation}
where on the right hand side we have three superoperators acting on
the RDO. The first of those is the Liouville superoperator associated
with the corresponding system Hamiltonian (Eq. \eqref{eq:sys_ham})
by the relation ${\cal L}_{S}\bullet\equiv\frac{1}{\hbar}\left[H_{S},\bullet\right]$.
The square brackets denote a commutator as usual, and $\hbar$ is
the reduced Plank's constant (from now on we set it to unity for convenience).
This part of the equation governs the coherent evolution of an isolated
system of electronic states. The second term represents the general
propagation scheme for the electronic system due to the interaction
with the bath DOF as detailed below. The last term is introduced to
treat the decay of excitons due to the nonradiative transitions from
the excited states to the ground state.

\subsection{Redfield Relaxation Scheme\label{sub:Redfield-Relaxation-Scheme}}

One of the most popular methods of treating an excitonic system coupled
to a bath are the Redfield equations. They follow from the Liouville-von
Neumann equation under the assumption that the system-bath interaction
is sufficiently weak to perform a perturbative expansion. In the second
order approximation in the interaction Hamiltonian $H_{SB}$ one gets
the so-called Quantum Master Equation \citep{May2004}. Then, performing
the Markov approximation in the excitonic basis we get the dissipative
part of Eq. \eqref{eq:general_eq} reading:

\begin{equation}
(D[\rho])_{\mu\nu}=-\sum_{\mu'\nu'}{\cal R}_{\mu\nu,\mu'\nu'}\rho_{\mu'\nu'}.\label{eq:Redfield_eq}
\end{equation}
The tetradic relaxation matrix ${\cal R}_{\mu\nu,\mu'\nu'}(t)$ is
the Redfield tensor and it is most conveniently given as follows:

\begin{align}
{\cal R}_{\mu\nu,\mu'\nu'}(t)\equiv & \delta_{\nu\nu'}\sum_{\varepsilon}\Gamma_{\mu\varepsilon,\varepsilon\mu'}(t)+\delta_{\mu\mu'}\sum_{\varepsilon}\Gamma_{\nu\varepsilon,\varepsilon\nu'}^{*}(t)\label{eq:Redfield tensor}\\
 & -\Gamma_{\nu'\nu,\mu\mu'}(t)-\Gamma_{\mu'\mu,\nu\nu'}^{*}(t).\nonumber 
\end{align}
Here, $\delta_{\nu\nu'}$ is the Kronecker's delta, and $\Gamma$'s
are certain damping matrices defined as follows:

\begin{equation}
\Gamma_{\mu\nu,\mu'\nu'}(t)\equiv\sum_{ij}\langle\mu|K_{i}|\nu\rangle\langle\mu'|K_{j}|\nu'\rangle\int_{0}^{t}\textrm{d}\tau C_{ij}(\tau)\textrm{e}^{\textrm{i}\omega_{\nu'\mu'}\tau}.\label{eq:Gamma_matrx}
\end{equation}
The matrix elements $\langle\mu|K_{i}|\nu\rangle$ represent the basis
transformation from molecular states (sites) to the delocalized eigenstates,
and $C_{ij}(t)$ is the energy gap correlation function in the site
basis

\begin{equation}
C_{ij}(t)\equiv\langle\Delta V_{i}(t)\Delta V_{j}(0)\rangle_{q}.\label{eq:Corr_func}
\end{equation}
The operator $\Delta V_{i}(t)=U_{B}^{\dagger}(t)\Delta V_{i}U_{B}(t)$,
where $U_{B}(t)$ is the bath evolution operator, represents the evolution
of the energy gap driven by the bath Hamiltonian $H_{B}$. In the
following we make an assumption that the energy gap fluctuations at
different sites are uncorrelated, i.e. $C_{ij}(t)=\delta_{ij}C_{i}(t)$.

Equations \eqref{eq:general_eq} can be solved once we have an explicit
form of the energy gap correlation functions (Eq. \eqref{eq:Corr_func})
$C(t)$ or spectral densities $C^{\prime\prime}(\omega)$. The two
quantities are related by

\begin{equation}
C_{i}(t)=\int_{0}^{\infty}\textrm{d}\omega\left\{ (1+n(\omega))\textrm{e}^{-\textrm{i}\omega t}+n(\omega)\textrm{e}^{\textrm{i}\omega t}\right\} C_{i}^{\prime\prime}(\omega),\label{eq:corr_f vs spectr_d}
\end{equation}
where $n(\omega)$ denotes the Bose-Einstein distribution function:

\begin{equation}
n(\omega)\equiv\frac{1}{\exp(\omega\beta)-1}.\label{eq:Bose-Einstein}
\end{equation}
A well explored model of the bath corresponding to a continuous spectrum
of the phonons is the so-called Debye spectral density \citep{M.G.Mueller2010,Rui-XueXu2009,mukbook}:

\begin{equation}
C_{i}^{\prime\prime}(\omega)=2\lambda_{i}\frac{\omega\gamma}{\omega^{2}+\gamma^{2}},\label{eq:BO spectral density}
\end{equation}
where the parameter $\gamma$ characterizes the timescale of the dissipation
of the reorganization energy, and the relation $\int_{0}^{\infty}\textrm{d}\omega C_{i}^{\prime\prime}(\omega)/\omega\pi=\lambda_{i}$
holds. Hence we can characterize each monomer in our system by the
reorganization energy $\lambda_{i}$ defined in Eq. \eqref{eq:reorg. energy},
which is not just the energy shift in Eq. \eqref{eq:sys_ham}, but
also the measure of the system-bath coupling. The parameter $\gamma$,
which is purely the bath property, is identical for both monomers
under the assumption of identical baths.

At this stage the Redfield equations can already be used in calculations,
however, under the conditions justifying the Markov approximation
the related electronic dynamics is much slower then the decay of the
bath correlation and the scheme can be simplified by shifting the
integration limit in Eq. \eqref{eq:Gamma_matrx} to $t\rightarrow\infty$.
This yields a time-independent Redfield tensor. The so-called secular
approximation \citep{May2004}, which decouples the evolution of populations
from that of the coherences, also applies under these conditions.
Formally, this approximation is realized by setting to zero elements
of ${\cal R}_{\mu\nu,\mu'\nu'}$ with the indices other than those
satisfying the conditions $\mu=\nu$ and $\mu'=\nu'$ or $\mu=\mu'$
and $\nu=\nu'$. In this paper both versions of the Redfield theory
- the full and the time-independent secular - are used for comparison.

\subsection{HEOM Relaxation Scheme\label{sub:HEOM-Relaxation-Scheme}}

Recently the hierarchical equations of motion approach has been introduced
for the description of the exciton dynamics (see, for instance, \citep{ishizaki:234111}).
Since it employs the cumulant expansion technique, it is exact if
the bath is Gaussian. This is the case for a bath of harmonic oscillators.
Unlike with the other methods, such as the Redfield equations, the
precise form of HEOM depends on the specific form of the bath correlation
function. Here we use an approximate HEOM theory termed the hierarchical
quantum master equation (HQME) \citep{Rui-XueXu2009}. It is based
on the following form of the correlation function, which is an approximate
expansion of Eq. \eqref{eq:corr_f vs spectr_d} with the Debye spectral
density defined by Eq. \eqref{eq:BO spectral density}:

\begin{equation}
C_{i}(t)=\left(\frac{2\lambda_{i}}{\beta}-\frac{\beta\lambda_{i}\gamma^{2}}{6}\right)\mathrm{\textrm{e}}^{-\gamma t}-\mathrm{i}\lambda_{i}\gamma\mathrm{\textrm{e}}^{-\gamma t}+\frac{\lambda_{i}\gamma\beta}{3}\delta\left(t\right);\label{eq:corr_f expanded}
\end{equation}
where $\delta\left(t\right)$ is the Dirac delta function.

In this method we replace the RDO $\hat{\rho}$ in Eq. \eqref{eq:general_eq}
with a hierarchy of operators $\hat{\rho}\rightarrow\hat{\rho}_{\mathbf{n}}$
indexed by $\mathbf{n}\equiv\left(n_{1},n_{2},\ldots,n_{M}\right)$,
where $n_{i}$'s are non-negative integers and $M$ is the number
of states in the system. In this set, the operator denoted by $\rho_{\mathbf{0}}$,
where $\mathbf{0}\equiv\left(0,0,\ldots,0\right)$, is the RDO. Other
operators are auxiliary and they contain information about the system-bath
correlations. The dissipation term ${\cal D}[\hat{\rho}]$ reads within
the HQME as:

\begin{alignat}{1}
{\cal D}[\hat{\rho}_{\mathbf{n}}] & =-\sum_{i=1}^{M}n_{i}\gamma\rho_{\mathbf{n}}-\sum_{i=1}^{M}\delta\mathcal{R}_{i}\rho_{\mathbf{n}}\nonumber \\
 & -\mathrm{i}\sum_{i=1}^{M}\mathcal{B}_{i}\rho_{\mathbf{n_{\mathbf{\mathit{i}}}^{+}}}-\mathrm{i}\sum_{i=1}^{M}n_{i}\mathcal{A}_{i}\rho_{\mathbf{n_{\mathit{i}}^{-}}},\label{eq:HQMA main}
\end{alignat}
where we have defined the following superoperators:

\[
\delta\mathcal{R}_{i}\bullet=\frac{\lambda_{i}\gamma\beta}{6}\mathcal{B}_{i}\mathcal{B}_{i}\bullet,
\]

\[
\mathcal{A}_{i}\bullet=\left(\frac{2\lambda_{i}}{\beta}-\frac{\lambda_{i}\gamma^{2}\beta}{6}\right)\mathcal{B}_{i}\bullet-\mathrm{i}\lambda_{i}\gamma\left\{ K_{i},\bullet\right\} ,
\]
and $\mathcal{B}_{i}\bullet=\left[K_{i},\bullet\right]$. The curly
brackets denote the anticommutator, and $\mathbf{n}_{i}^{\pm}\equiv\left(n_{1},n_{2},\ldots,n_{i}\pm1,\ldots,n_{M}\right)$.
The operators $\hat{\rho}_{\mathbf{n}}$ with certain $\mathbf{n}$
such that $N=\sum_{i}n_{i}$, define a \textquotedbl{}tier N\textquotedbl{}
of operators. We can see that within Eq. \eqref{eq:HQMA main} each
tier $N$ becomes coupled to neighboring tiers $N\pm1$. Formally,
the hierarchy in HQME continues to infinity. In practice, however,
one has to truncate the hierarchy by setting all auxiliary density
operators from the tiers $N>N_{trunc}$ to zero. The cutoff $N_{trunc}$
is defined by the relative values of the model parameters. In calculations
it can simply be chosen to ensure the convergence of the solution
for $\rho_{\mathbf{0}}$.

\subsection{Exciton Decay\label{sub:Exciton-Decay}}

In contrast to the second term in Eq. \eqref{eq:general_eq}, which
is described by relaxation theories of previous subsections, we include
decay of excitons in a phenomenological way by taking the experimentally
estimated rates for population decay processes into account. Therefore,
the ${\cal K}$ superoperator elements are phenomenologically defined
in the site basis. We assume that the only non-zero elements of this
superoperator are those that describe the decay of populations and
the corresponding decay of coherences, i.e., we assume the secular
approximation. The elements of the superoperator thus read:

\begin{equation}
{\cal K}_{ij,kl}=-\frac{\kappa_{i}+\kappa_{j}}{2}\delta_{ik}\delta_{jl},\label{eq:relaxation tensor}
\end{equation}
where $\kappa_{i}$ denotes the relaxation rate of the population
of the \textit{i}th site. This way we have only two types of elements:
the population relaxation (${\cal K}_{ii,ii}$) and the coherence
decay due to corresponding population relaxation (${\cal K}_{ij,ij}$).

In order to use the ${\cal K}$ tensor with the Redfield equations,
it needs to be transformed into the excitonic basis. The transformation
matrix for a superoperator is given in the Appendix. In the case of
a dimer we obtain simple expressions with the mixing angle once again.
For the population relaxation the tensor elements read:

\begin{align}
{\cal K}_{\alpha\alpha,\alpha\alpha}= & {\cal K}_{aa,aa}\cos^{2}\theta+{\cal K}_{bb,bb}\sin^{2}\theta;\nonumber \\
{\cal K}_{\beta\beta,\beta\beta}= & {\cal K}_{aa,aa}\sin^{2}\theta+{\cal K}_{bb,bb}\cos^{2}\theta.\label{eq:excitonic rates}
\end{align}

\section{Results}

\begin{figure}
\centering{}\includegraphics[scale=0.55]{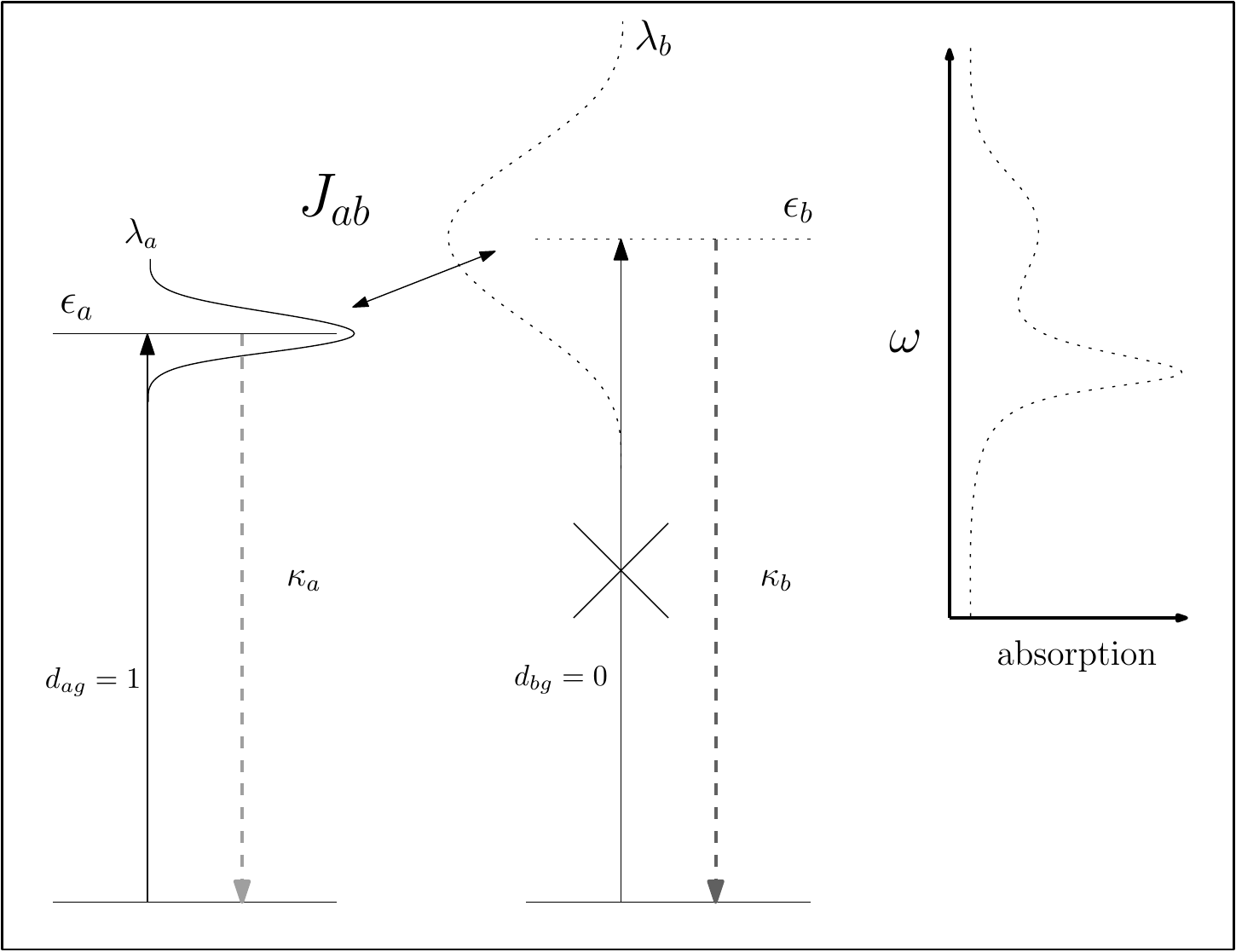}\caption{Schematic representation of the constituent monomers of the heterodimer.
The excited states of the monomers are indicated as $\epsilon_{a}$
and $\epsilon_{b}$, correspondingly, the resonance interaction between
the monomers is indicated as $J_{ab}$. The optical transition in
the $a$-monomer is allowed, while it is forbidden in the $b$-monomer
as indicated by the vertical black arrows. The inset shows a hypothetical
absorption spectrum of such system. The position of $\epsilon_{b}$
might be above or below $\epsilon_{a}$ and therefore is shown by
dotted line; reorganization energies are shown to correspond to the
homogeneous linewidths of the absorption spectrum.}
\label{fig:Heterodimer}
\end{figure}

We set up our heterodimer to have some typical properties of photosynthetic
Chl\textendash{}Car aggregates. As discussed above, such a dimer might
be relevant to the quenching in photosynthesis. Therefore, one of
the constituent monomers is characterized by an optically dark and
extremely short-lived excited state. The other monomer has a strong
transition dipole moment and a long-lived excited state. Their homogeneous
broadenings are different as well. This situation is schematically
depicted in Fig. \ref{fig:Heterodimer}.

The constant parameters are the following: $T=300\, K$ (which corresponds
to $k_{B}T\approx208\, cm^{-1}$), $J_{ab}=100\, cm^{-1}$, $\gamma^{-1}=100\, fs$,
dipole moments $d_{ag}=1,\: d_{bg}=0$ in accord with Fig. \ref{fig:Heterodimer}.
The tunable parameters are the reorganization energies $\lambda_{a},\:\lambda_{b}$,
given as four combinations of the values $30\, cm^{-1}$ and $150\, cm^{-1}$,
and the gap between the site excitation energies with values $\pm100\, cm^{-1}$
(\textquotedbl{}$+$\textquotedbl{} corresponds to the bright state
being above the dark one; \textquotedbl{}$-$\textquotedbl{} corresponds
to the opposite situation).

We will consider optical excitation as a trigger of the relaxation
dynamics. The initial condition for the evolution of the RDO corresponding
to a resonance excitation of the optically allowed state is then given
by $\rho_{\mu\mu}(0)=|d_{\mu g}|^{2}$, where $d_{\mu g}$ is the
matrix element of the transition dipole moment operator in the excitonic
basis.

\begin{figure}
\begin{centering}
\includegraphics[scale=0.52]{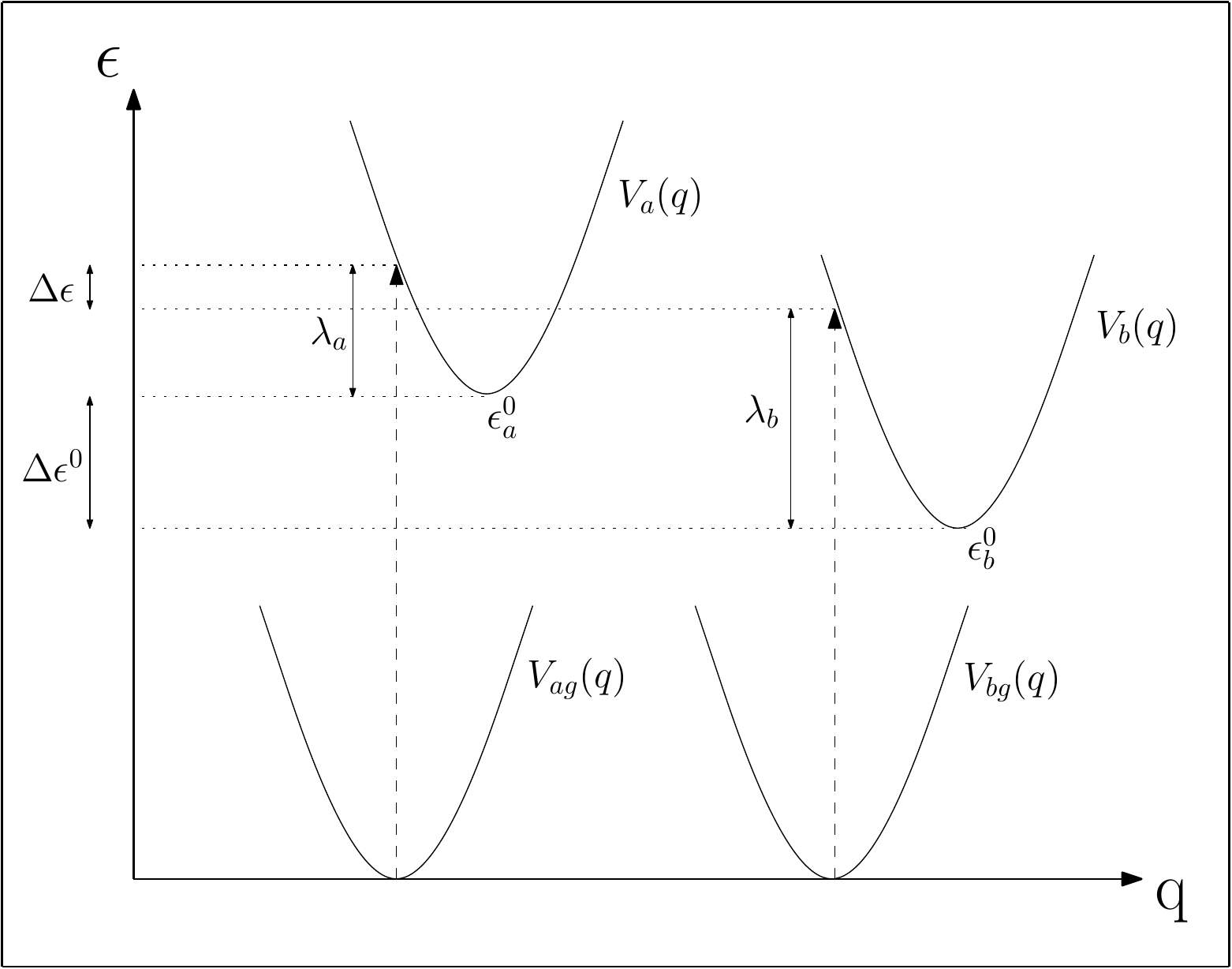}
\par\end{centering}

\centering{}\caption{Definition of the energy gap in the case of different reorganization
energies. The dashed vertical arrows indicate the Franck-Condon transitions.
Clearly $\lambda_{a}\neq\lambda_{b}$ corresponds to $\Delta\epsilon^{0}\neq\Delta\epsilon$.}
\label{fig:definition of gap}
\end{figure}

The main parameter characterizing a heterodimer is the difference
of site energies, but as can be seen in Fig. \ref{fig:definition of gap},
upon the presence of the bath, in the case of $\lambda_{a}\neq\lambda_{b}$,
the definition of the energy gap becomes somewhat ambiguous: we can
define it either as $\Delta\epsilon=\epsilon_{a}-\epsilon_{b}$ or
as $\Delta\epsilon^{0}=\epsilon_{a}^{0}-\epsilon_{b}^{0}$ (even though
the relation $\Delta\epsilon=\Delta\epsilon^{0}+\lambda_{a}-\lambda_{b}$
holds). Application of these different definitions of a dimeric energy
gap has a distinct influence on the modeled dynamics.

\subsection{Population Kinetics\label{sub:Population-Kinetics}}

\begin{figure*}
\begin{centering}
a)\includegraphics[scale=0.3]{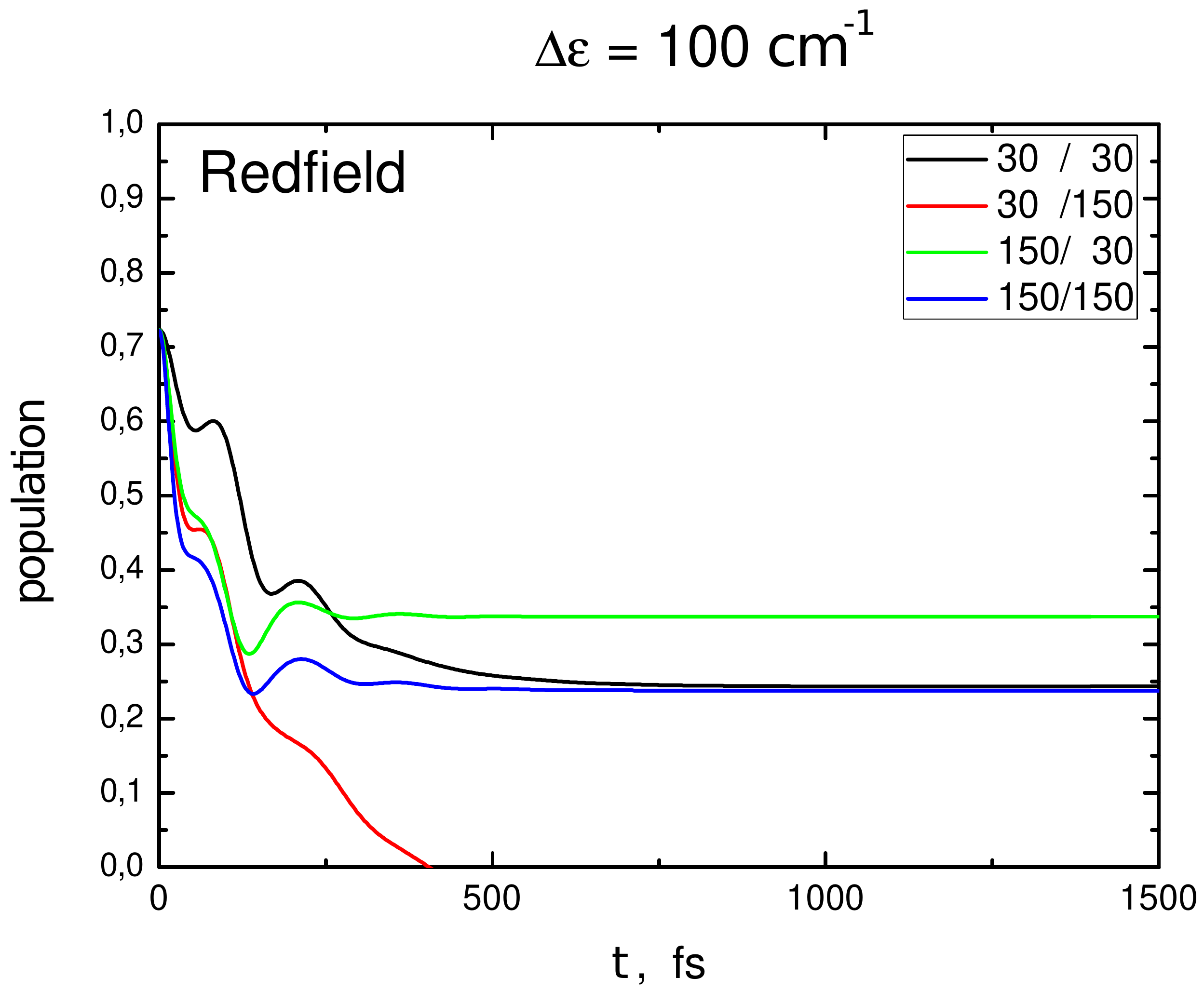}b)\includegraphics[scale=0.3]{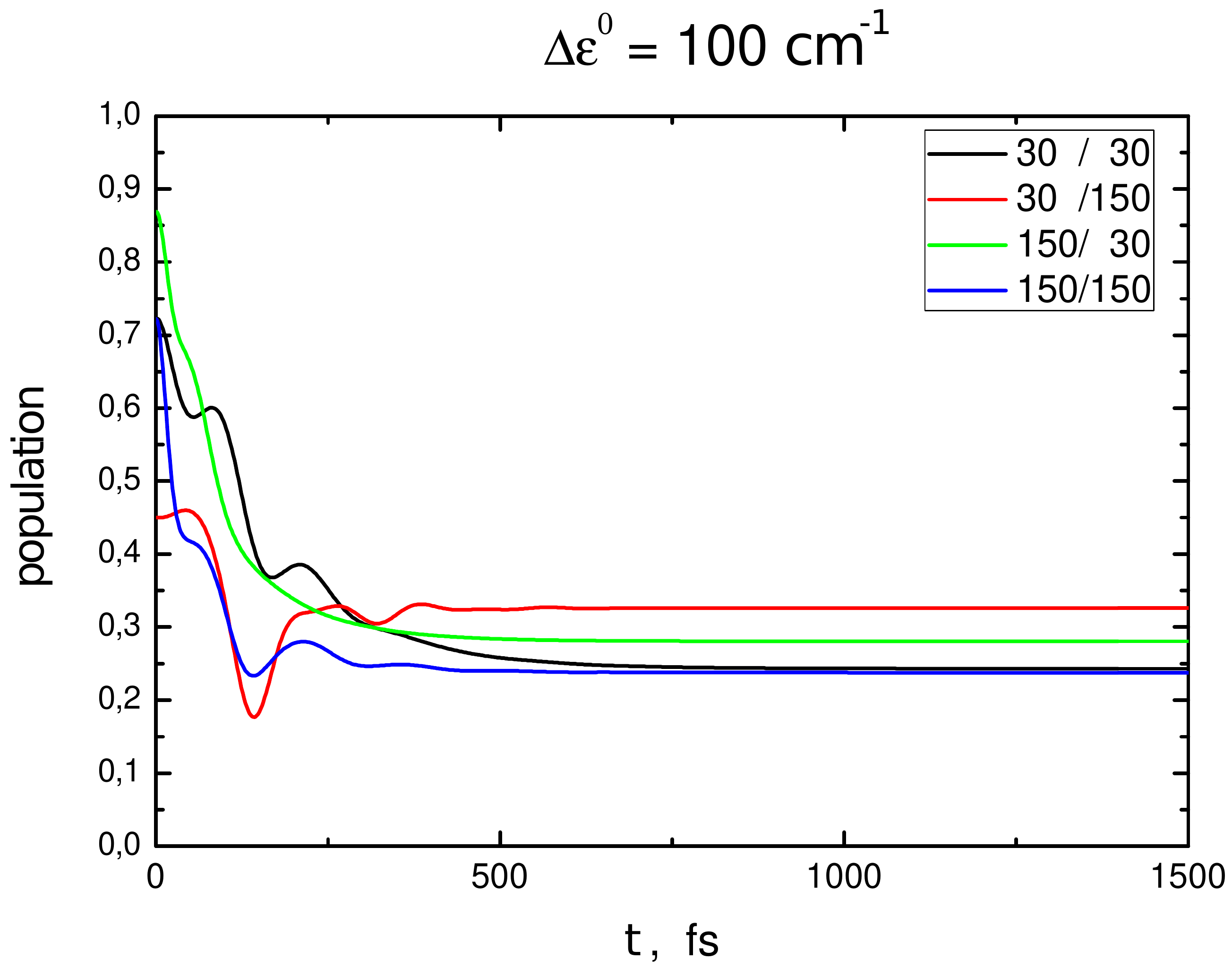}
\par\end{centering}

\begin{centering}
c)\includegraphics[scale=0.3]{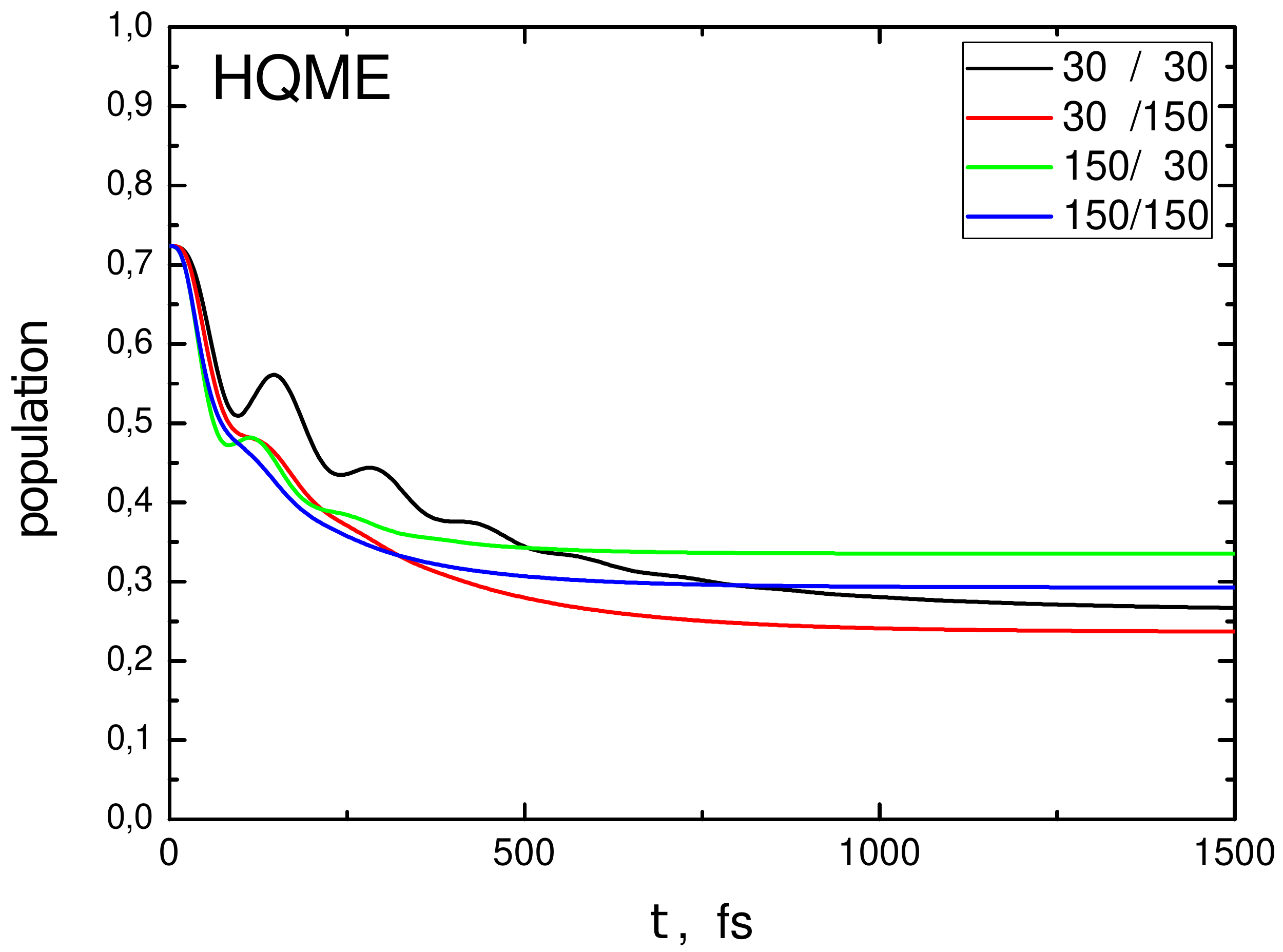}d)\includegraphics[scale=0.3]{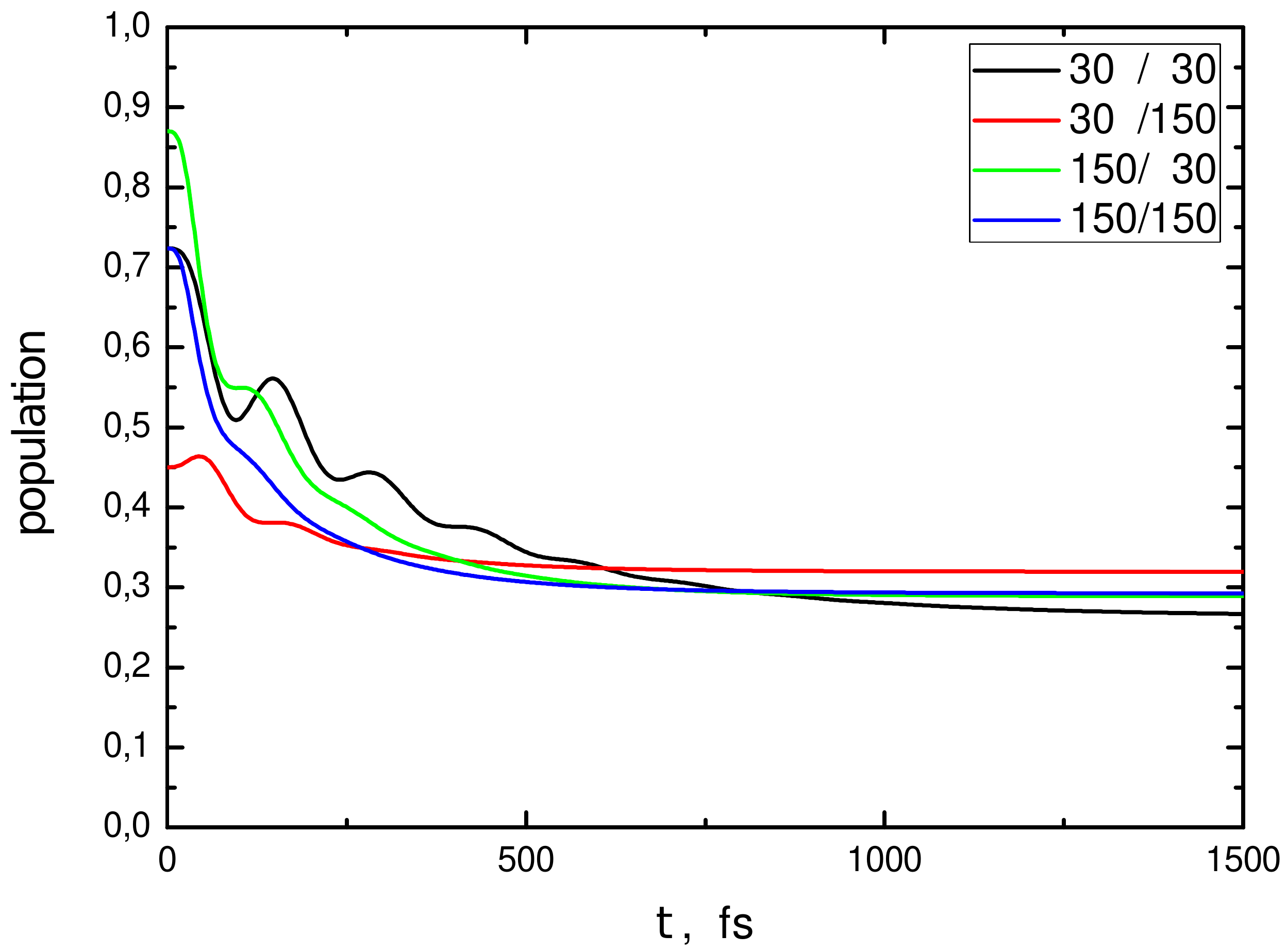}
\par\end{centering}

\begin{centering}
e)\includegraphics[scale=0.3]{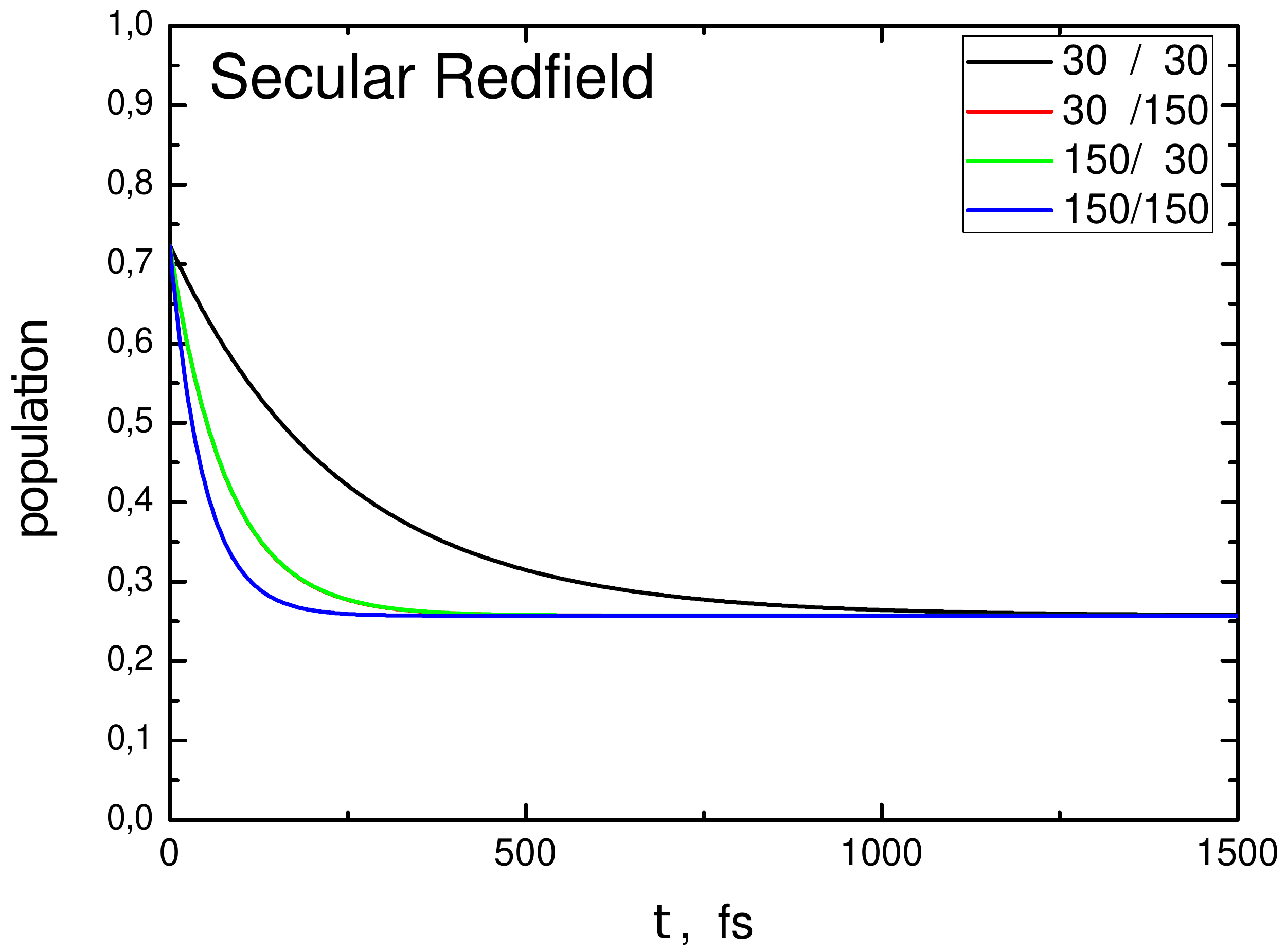}f)\includegraphics[scale=0.3]{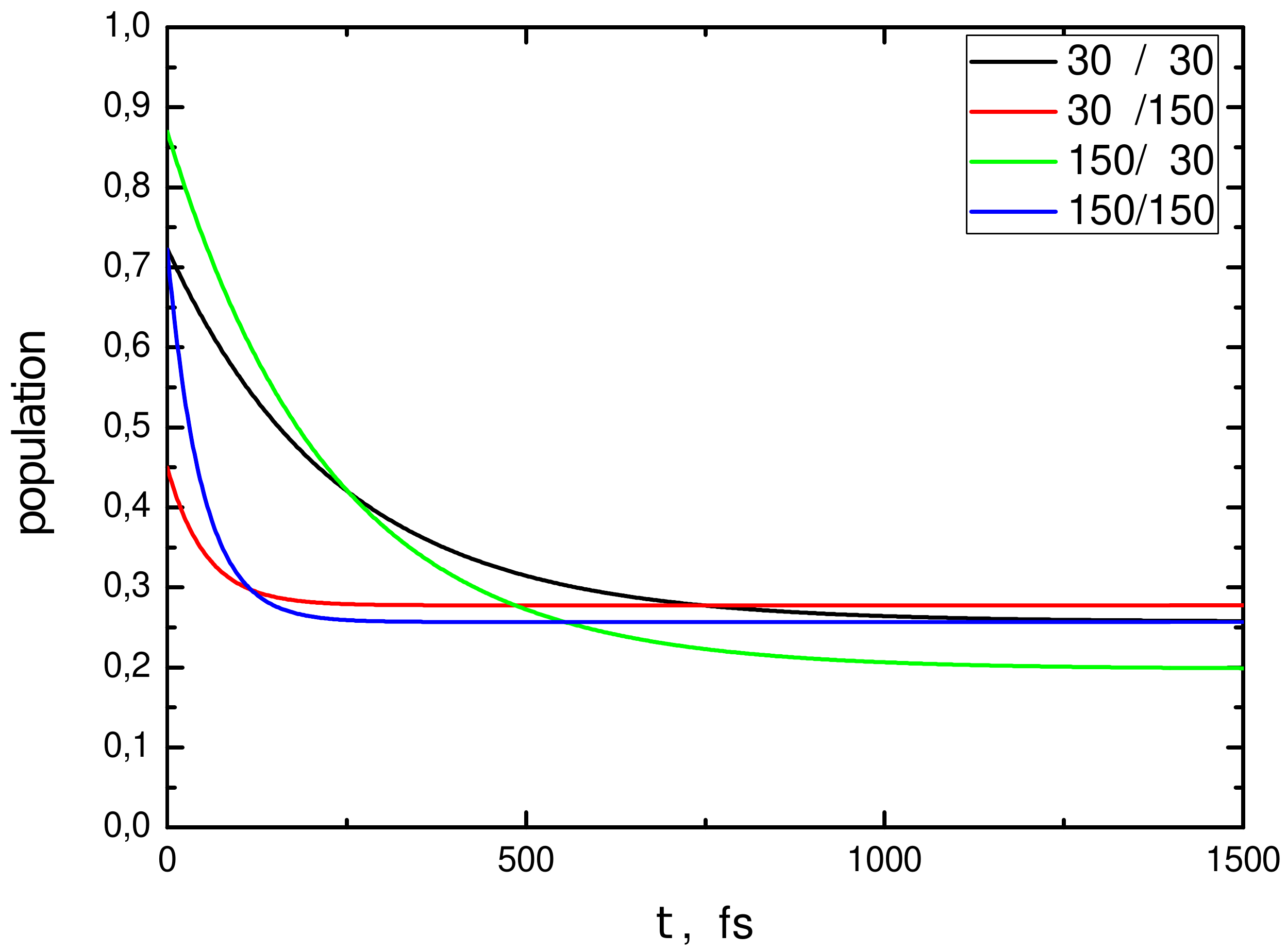}
\par\end{centering}

\caption{Evolution of the higher excitonic state population. The insets show
the combination of reorganization energies given in the form $\lambda_{a}/\lambda_{b}$
($cm^{-1}/cm^{-1}$). The left column corresponds to $\Delta\epsilon=+100\, cm^{-1},$
the right one - to $\Delta\epsilon^{0}=+100\, cm^{-1}$. a) and b),
c) and d), e) and f) correspond to the full Redfield scheme, the HQME
and the secular Redfield equations, respectively.}
\label{fig:main evolutions}
\end{figure*}

We first study the dynamics of the excited state in the absence of
relaxation to the ground state, i.e. when ${\cal K}=0$. In Fig. \ref{fig:main evolutions}
we show the time evolution of the higher excitonic state population
modeled by the Redfield equations (a and b), the HQME (c and d) and
the secular Redfield equations (e and f) using four combinations of
the reorganization energies. The left column corresponds to $\Delta\epsilon=+100\, cm^{-1}$
as a fixed energy gap, while the right column corresponds to $\Delta\epsilon^{0}=+100\, cm^{-1}$
being fixed. The equal reorganization energies (black and blue curves)
serve as a good starting point for comparison of the methods. As we
can see, the initial stages of the Redfield and HQME solutions ($\sim100\, fs$)
look very similar while the rates of the thermalization and the frequencies
of the coherent oscillations are different (the coherent oscillations
are more dramatic but shorter-lived in the Redfield solution). Moreover,
different methods give us different equilibrium population values,
and within the HQME solutions the latter do not coincide for two pairs
of identical $\lambda$'s.

The case of different reorganization energies (red and green curves)
is quite nontrivial \citep{T.Mancal2006}: the results for both energy
gap definitions obtained by all three methods are considerably different
in the long-time limit. In the right-hand column the initial values
are scattered simply because fixing $\Delta\epsilon^{0}$ with different
$\lambda$'s gives us different $\Delta\epsilon$ used in the definition
of the excitonic basis. Moreover, the red curve in Fig. \ref{fig:main evolutions}a
reveals the well-known problem of the Redfield scheme, namely, that
it does not guarantee positivity. The secular Redfield solutions do
not suffer from this problem, and in this case all four $\lambda$
combinations give relaxation to the same equilibrium position.

\begin{figure}
\begin{centering}
\includegraphics[scale=0.35]{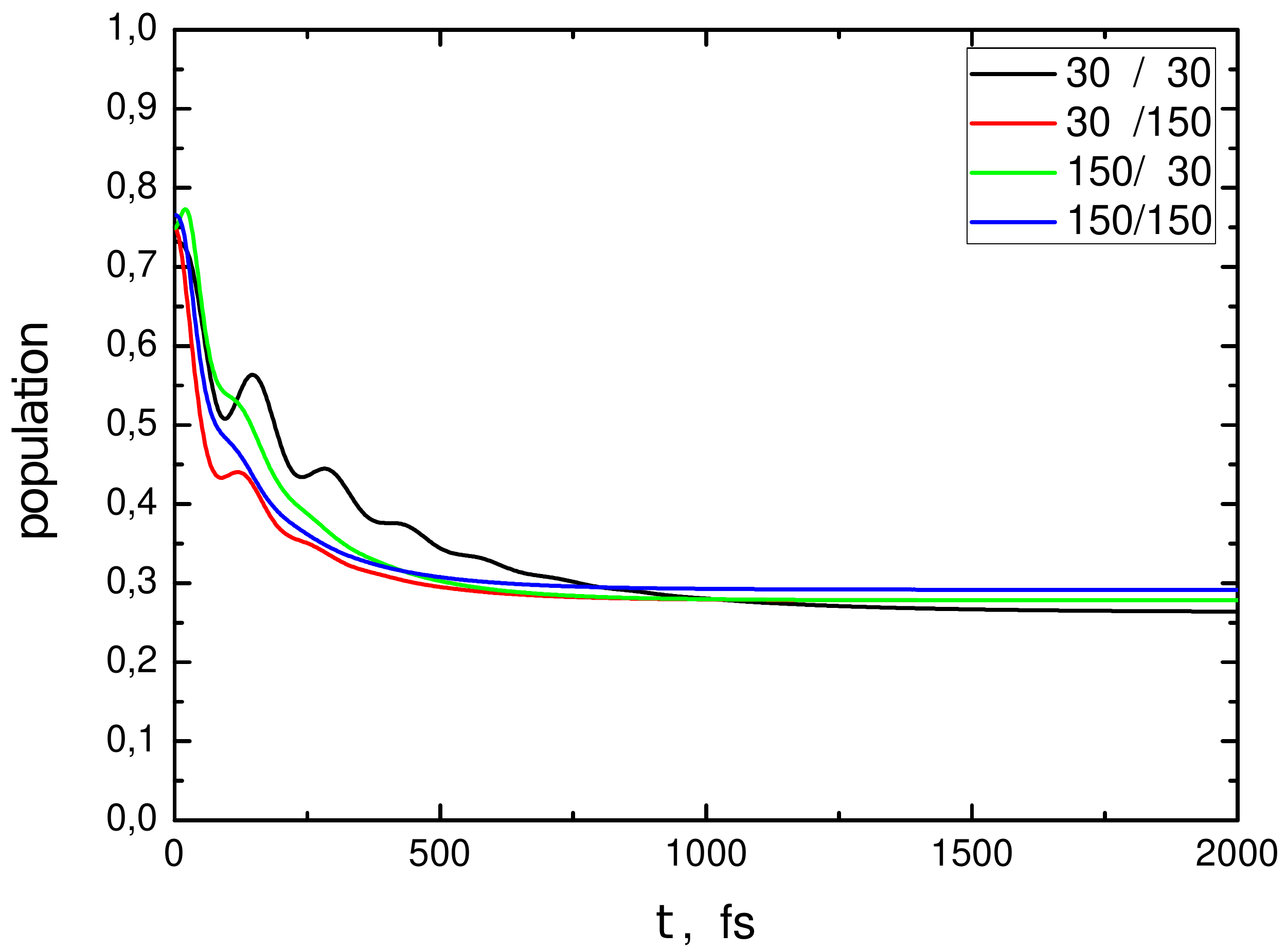}
\par\end{centering}

\caption{The HQME solution of the $\Delta\epsilon^{0}=100\, cm^{-1}$ configuration
in the preferred basis. The inset shows the combination of reorganization
energies given in the form $\lambda_{a}/\lambda_{b}$ ($cm^{-1}/cm^{-1}$).}

\label{fig:polaron basis}
\end{figure}

So far we have treated the excitonic states defined by Eq. \eqref{eq:exc_expa}
as the eigenstates of the system. However, HQME solutions in the excitonic
basis, as given above, exhibit non-vanishing steady state coherences.
The presence of the steady state coherences can be interpreted as
a reflection of renormalization of the system eigenbasis \citep{olsina-2010}
taking place in the course of time. By virtue of the exact treatment
of the system-bath interaction the HQME solutions are basis-independent,
which allows us to identify the so-called preferred basis (or \textquotedbl{}global
basis\textquotedbl{} \citep{Gelzinis2011}) in which the density matrix
is diagonal in the long-time limit (see, e.g. \citep{Schlosshauer2007}).
This is done by diagonalizing the stationary part of the solution
in the excitonic basis. Representing the RDO in the preferred basis
for the fixed $\Delta\epsilon$ configuration has the effect of slightly
shifting the equilibrium population values. Their order, however,
remains the same as in the excitonic basis. Meanwhile for the fixed
$\Delta\epsilon^{0}$ configuration (Fig. \ref{fig:polaron basis}),
we have not just quantitative but also qualitative changes in comparison
with Fig. \ref{fig:main evolutions}d. Evidently, the solutions for
$\lambda_{1}\neq\lambda_{2}$ configurations now converge into the
same steady state value which is in-between the ones for equal reorganization
energies.

\subsection{Relaxation to the Ground State\label{sub:Relaxation-to-the}}

\begin{figure}
\begin{centering}
\includegraphics[scale=0.34]{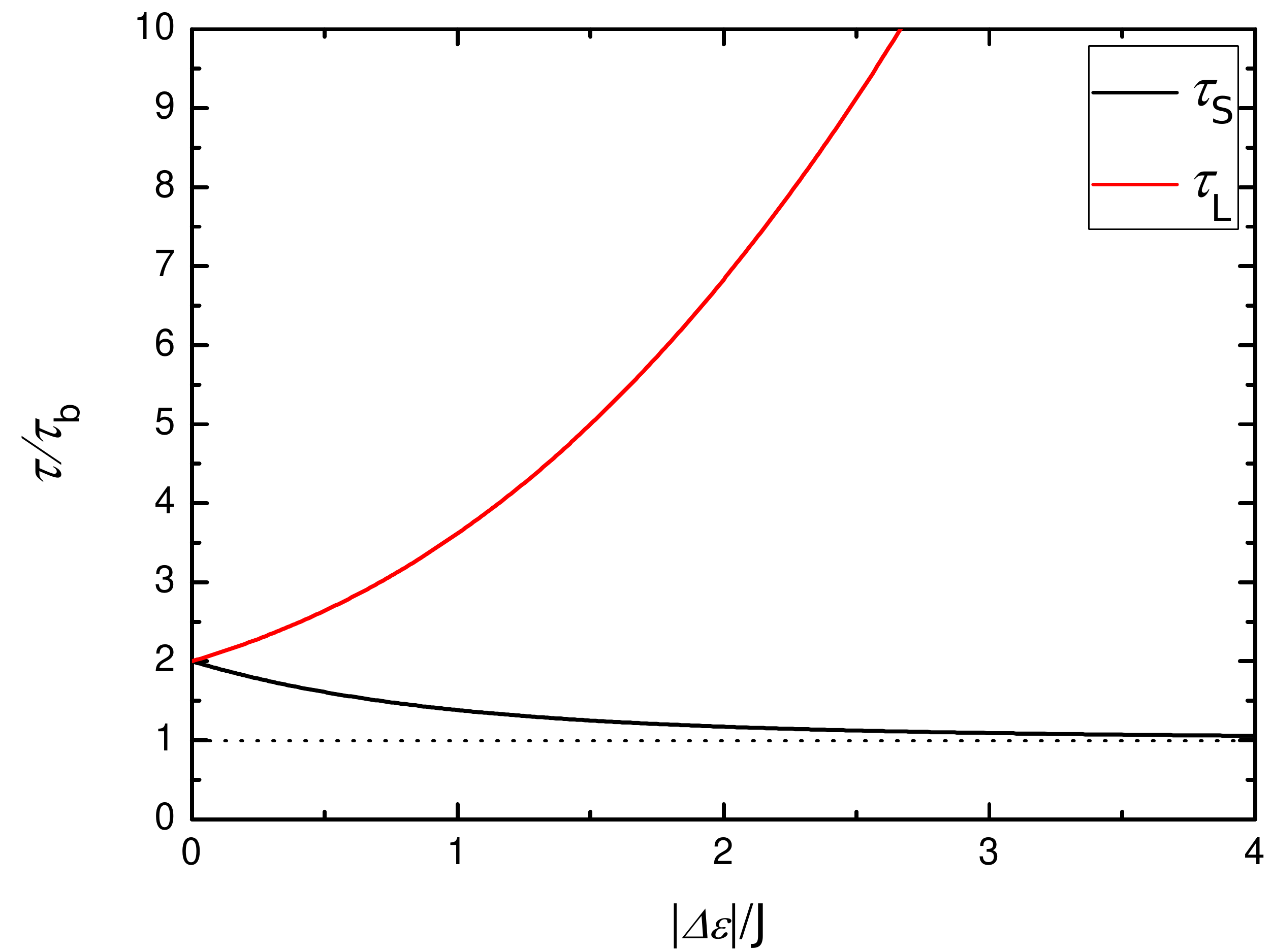}
\par\end{centering}

\caption{Excitonic mixing of lifetimes. Indices \textquotedbl{}S\textquotedbl{}
and \textquotedbl{}L\textquotedbl{} denote the short-lived and the
long-lived states accordingly. The energy gap $\Delta\epsilon$ is
normalized to the resonance coupling, and the lifetimes are normalized
to the lifetime of the short-lived state in the absence of resonance
coupling ($\tau_{b}$).\label{fig:Mixing-of-lifetimes}}
\end{figure}

We next include the relaxation to the ground state. The relaxation
tensor of the equation of motion for the RDO is given by Eq. \eqref{eq:relaxation tensor}.
We are interested in the influence of the resonance coupling on the
lifetimes of the dimer eigenstates $\tau_{i}=\kappa_{i}^{-1}$. For
our heterodimer we assume that in the absence of coupling the excitation
of one state relaxes considerably faster than the other ($\tau_{b}\ll\tau_{a}$
or ${\cal K}_{aa,aa}\ll{\cal K}_{bb,bb}$). Therefore we can simplify
Eqs. \eqref{eq:excitonic rates} to

\begin{align}
{\cal K}_{\alpha\alpha,\alpha\alpha}= & {\cal K}_{bb,bb}\sin^{2}\theta;\label{eq:excitonic rates approx}\\
{\cal K}_{\beta\beta,\beta\beta}= & {\cal K}_{bb,bb}\cos^{2}\theta.\nonumber 
\end{align}
Denoting the lifetimes of the short-lived state by $\tau_{S}\equiv\tau_{\beta}=1/{\cal {\cal K}}_{\beta\beta,\beta\beta}$
and the long-lived one by $\tau_{L}\equiv\tau_{\alpha}=1/{\cal {\cal K}}_{\alpha\alpha,\alpha\alpha}$
we present their dependence on the (site) energy gap normalized to
the resonance coupling in Fig. \ref{fig:Mixing-of-lifetimes}. The
lifetimes are normalized to the initial lifetime of the (original,
i.e. uncoupled) short-lived state $\tau_{b}$.

\begin{figure}
\begin{centering}
\includegraphics[scale=0.35]{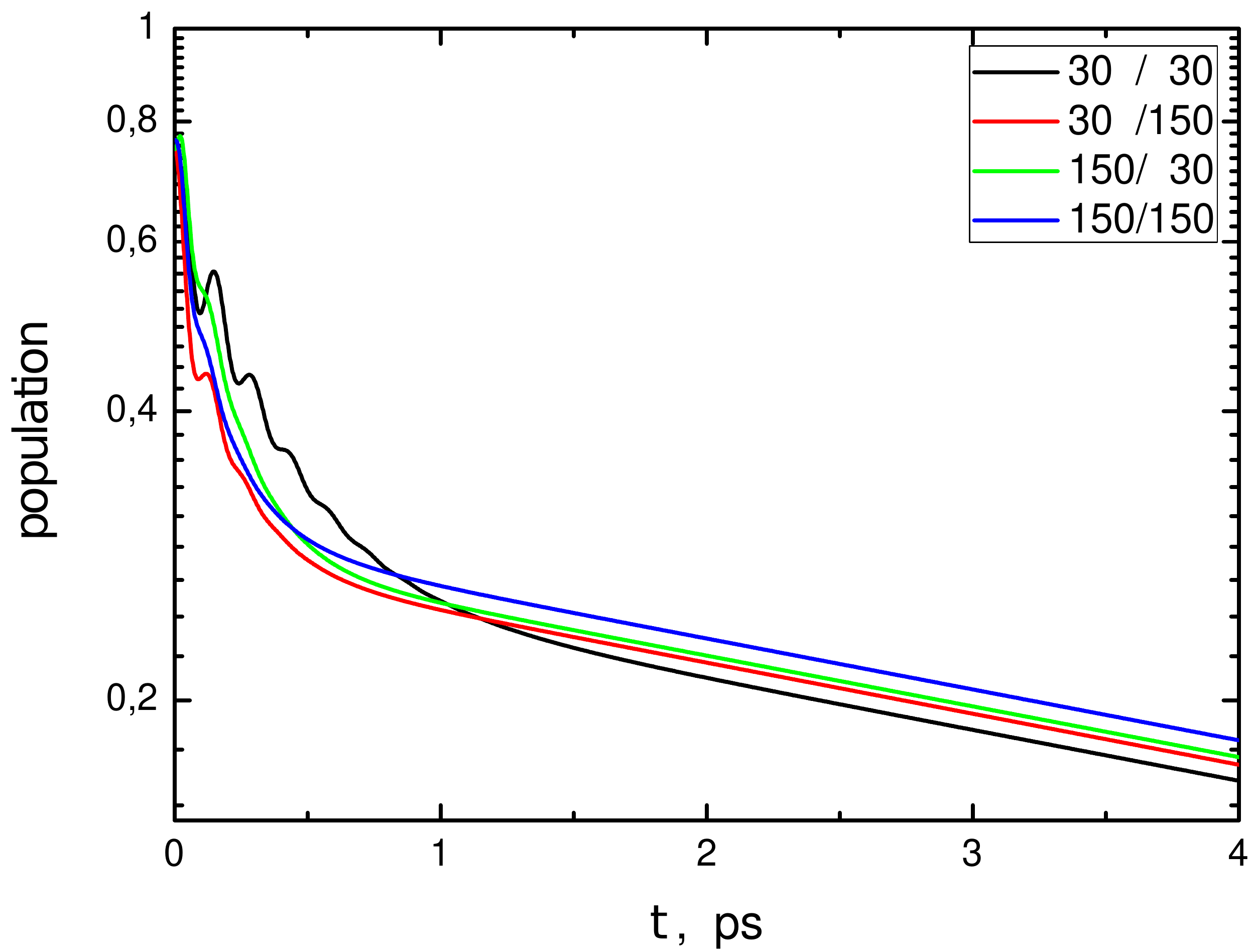}
\par\end{centering}

\caption{Solutions of the HQME with relaxation. The populations are given in
the logarithmic scale. The inset shows the combination of reorganization
energies given in the form $\lambda_{a}/\lambda_{b}$ ($cm^{-1}/cm^{-1}$).\label{fig:all 4 relaxations HQME}}
\end{figure}

Now let us consider the interplay between processes of thermalization
as described in the previous subsection and relaxation to the ground
state. The numerical results were obtained from the HQME with the
relaxation tensor and are presented in the preferred basis. The results
for a system with $\Delta\epsilon=100\, cm^{-1}$ and $\tau_{a}=2\, ns$,
$\tau_{b}=5\, ps$ are shown in Fig. \ref{fig:all 4 relaxations HQME}
(the corresponding \textquotedbl{}excitonic\textquotedbl{} life-times
yield $\tau_{\alpha}\approx21\, ps$ and $\tau_{\beta}\approx7\, ps$).
The populations are given in the logarithmic scale in order to reveal
the two-exponential nature of the process. Only the higher state population
evolutions are shown, because the lower state evolves with identical
rates. The result is similar to the one shown in Fig. \ref{fig:polaron basis}
except that instead of the steady state values we have now a decay
due to relaxation to the ground state, and the long-time population
values for the $\lambda_{a}\neq\lambda_{b}$ configurations (red and
green curves) are no longer identical. Obviously, the long-time decay
rates are identical in all four cases.

To demonstrate the influence of the energetic position of the short-lived
state, the evolutions for the case of $\Delta\epsilon=\pm100\, cm^{-1}$
and $\lambda_{a}=\lambda_{b}=150\, cm^{-1}$ are considered. For demonstration,
we provide an analytical approximation of the evolution while neglecting
the coherent oscillations. This can be achieved by using a simple
system of rate equations:

\begin{equation}
\begin{cases}
\dot{x}= & -(\kappa_{x}+k_{x\rightarrow y})\, x\ +\ k_{y\rightarrow x}\, y;\\
\dot{y}= & -(\kappa_{y}+k_{y\rightarrow x})\, y\ +\ k_{x\rightarrow y}\, x.
\end{cases}\label{eq:app C main system}
\end{equation}
Here, $x$ and $y$ are the populations (of the accordingly lower
and higher states), $\kappa_{x}$ and $\kappa_{y}$ are the population
relaxation rates from the tensor ${\cal K}$, and $k_{x\rightarrow y},\, k_{y\rightarrow x}$
are the effective thermalization rates obtained from the data fit
of the solution of the HQME without relaxation to the ground state.
Following this approach, the population of the higher state reads:

\begin{equation}
y(t)=C_{y1}e^{-\xi_{1}t}+C_{y2}e^{-\xi_{2}t},\label{eq:app C solution}
\end{equation}
where $\xi_{1},\:\xi_{2}$ are the eigenvalues of Eq. \eqref{eq:app C main system},
and $C$'s are the corresponding amplitudes. When $k_{x\rightarrow y},\: k_{y\rightarrow x}\gg\kappa_{x},\:\kappa_{y}$
the second-order small terms can be neglected thus giving:

\begin{equation}
\begin{cases}
\xi_{1}\approx & -\frac{\kappa_{x}+\kappa_{y}}{2}+\frac{(k_{x\rightarrow y}-k_{y\rightarrow x})(\kappa_{x}-\kappa_{y})}{2(k_{x\rightarrow y}+k_{y\rightarrow x})};\\
\xi_{2}\approx & -(k_{x\rightarrow y}+k_{y\rightarrow x}).
\end{cases}\label{eq:app C eigenvalues}
\end{equation}
This allows us to estimate the dynamics in the following way. First,
we determine quantities $C_{y1},\: C_{y2}$ and $\xi_{2}$ from the
HQME solution without ${\cal K}$ (hence $\xi_{1}=0$) by numerical
fitting (disregarding the oscillations). Then we derive the rates
$k_{x\rightarrow y}$ and $k_{y\rightarrow x}$ from the known value
of $\xi_{2}$ and the detailed balance condition. It is noteworthy
that one must use an effective energy gap instead of the purely excitonic
one if the reorganization energies are large. Finally, we calculate
$\kappa_{x}$ and $\kappa_{y}$ from the relaxation superoperator
${\cal K}$ in the preferred basis, and then construct the $\xi_{1}$
eigenvalue.

\begin{figure}
\begin{centering}
\includegraphics[scale=0.35]{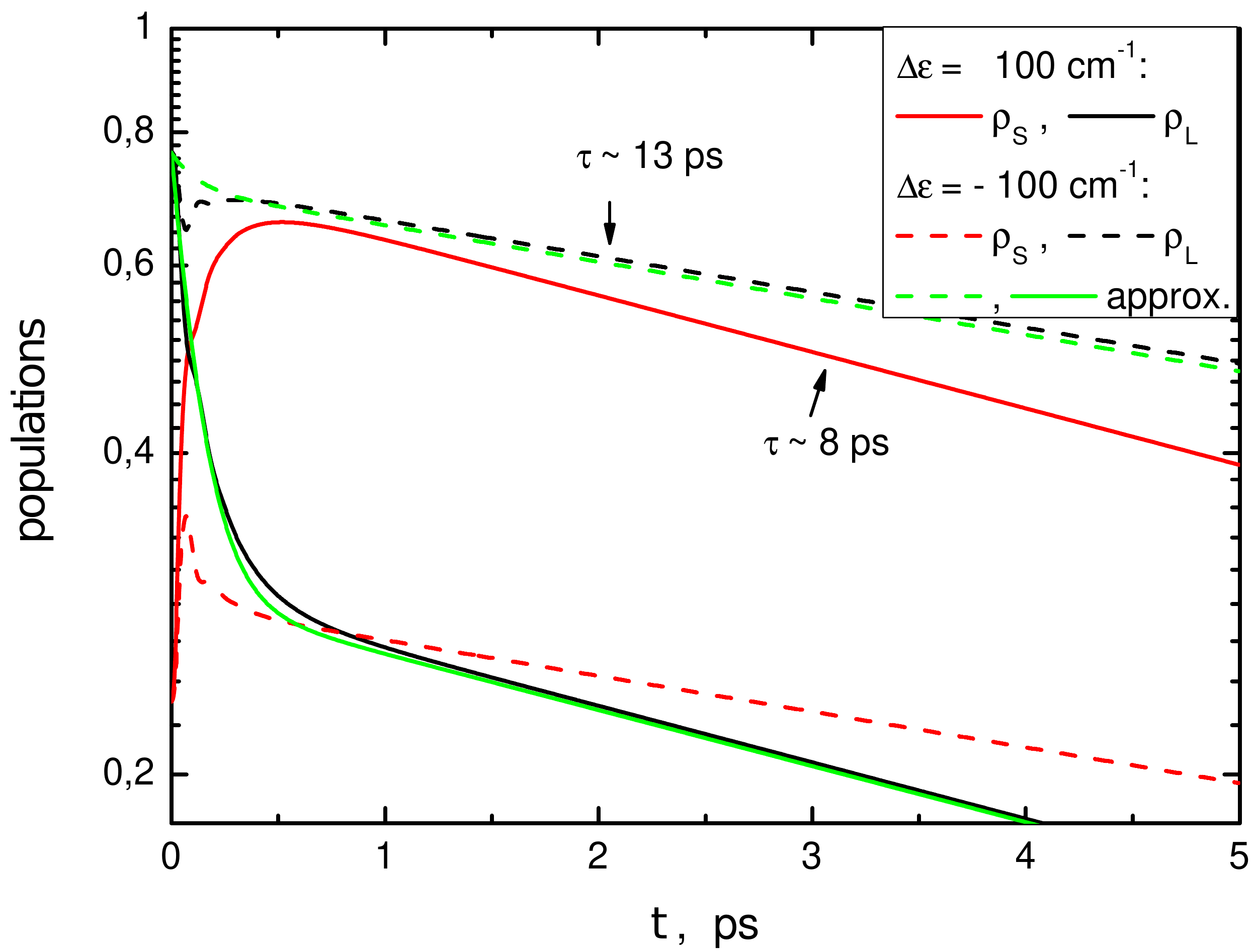}
\par\end{centering}

\caption{Solutions of the HQME with relaxation and approximations. The populations
are given in the logarithmic scale. Indices \textquotedbl{}S\textquotedbl{}
and \textquotedbl{}L\textquotedbl{} denote the short-lived and the
long-lived states accordingly.\label{fig:Solutions-of-HQME+R}}
\end{figure}

The results obtained are depicted in Fig. \ref{fig:Solutions-of-HQME+R}
using the logarithmic scale. The populations of the long-lived state
(in black) are shown together with their analytical approximations
(in green). In the case of $\Delta\epsilon=-100\, cm^{-1}$ (dashed
curves) the lower state is the long-lived one (hence, originating
from the bright electronic state), therefore the thermalization part
gives a minor effect. As can be seen, the actual rate of the relaxation
to the ground state, $\tau$, depends on which of the two states \textendash{}
the short-lived or the long-lived \textendash{} is the lower one in
the dimer.

The analysis given above provides an explanation of the long-time
decay rate ($\xi_{1}$) sensitivity to the position of the short-living
state. In particular, swapping the states results in swapping the
values of $\kappa_{x}$ and $\kappa_{y}$ in Eqs. \eqref{eq:app C main system}.
Therefore, the second term in the first equation from Eq. \eqref{eq:app C eigenvalues}
changes the sign, and that is the reason why the overall decay rate
is sensitive to which state \textendash{} the fast decaying or the
slowly decaying - is the lower one.

\section{Discussion\label{sec:Discussion}}

As follows from our calculations, the asymmetry in reorganization
energies allows for two non-equivalent definitions of the energy gap
in the one-exciton manifold. These two cases clearly have different
physical meanings. Since the energy $\epsilon$ refers to the Franck-Condon
transition region of the potential energy surface (see Fig. \ref{fig:definition of gap}),
$\Delta\epsilon$ corresponds to the distance between the peak positions
in the absorption spectra. Therefore we can loosely call it the \textquotedbl{}optical
energy difference\textquotedbl{}. This has a perfectly clear meaning
in the absorption spectroscopy, however, the thermal equilibrium does
not establish itself with respect to this energy gap. Should the resonance
coupling be simply perturbative, we could expect the thermalization
with respect to $\Delta\epsilon^{0}$ as the first approximation.
Since this gap is usually used in the F\"{o}rster resonance energy transfer
(FRET) theory \citep{valkunasbook,May2004} we shall call it the \textquotedbl{}F\"{o}rster
energy difference\textquotedbl{}.

\begin{figure}
\begin{centering}
\includegraphics[scale=0.53]{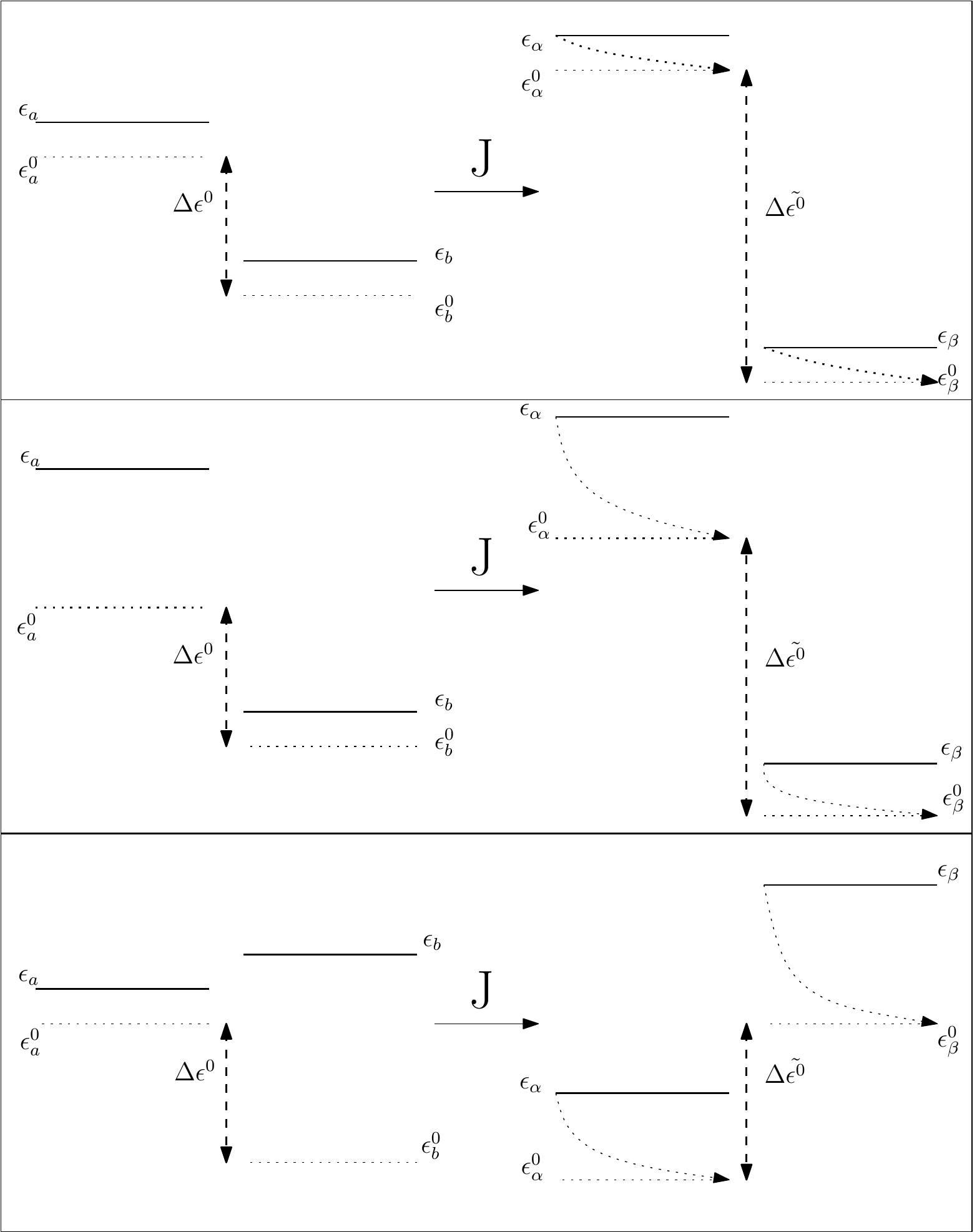}
\par\end{centering}

\caption{Schematic representation of the F\"{o}rster energy gap in the presence
of the resonance coupling. Monomeric units with identical $\Delta\epsilon^{0}$
are shown on the left. The dotted arrows on the right represent the
vibrational relaxation (and tilde is used to denote the excitonic
basis). The top row corresponds to $\lambda_{a}=\lambda_{b}$, the
middle and the bottom ones - to $\lambda_{a}>\lambda_{b}$ and $\lambda_{a}<\lambda_{b}$
accordingly. \label{fig:different Forster gaps}}
\end{figure}

For the fixed optical energy difference, in the case of $\lambda_{a}=\lambda_{b}$
an equal amount of energy associated with each state is dissipated
during the thermalization, and therefore the same equilibrium values
are reached. In the case of $\lambda_{a}\neq\lambda_{b}$, the vibrational
relaxation shifts the initial energies by a different amount ($\Delta\epsilon^{0}\neq\Delta\epsilon$).
Hence, according to the Redfield relaxation, in Fig. \ref{fig:main evolutions}a
the energy gap between the states is effectively decreased (green
curve) or increased (red curve) (here, the increase of the energy
gap is seemingly the reason for the failure of the Redfield scheme).
The case of fixed F\"{o}rster energy gap (Fig. \ref{fig:main evolutions}b)
demonstrates the excitonic mixing of reorganization energies even
better in the sense that for uncoupled monomers the vibrational relaxation
would yield identical $\Delta\epsilon^{0}$ regardless of the combination
of $\lambda$'s. The actual situation is schematically shown in Fig.
\ref{fig:different Forster gaps}. An interesting situation arises
in the case of $\lambda_{a}<\lambda_{b}$ (Fig. \ref{fig:main evolutions}b,
red curve), because the two states are swapped in comparison with
the other combinations (both initially and in the long-time limit).
The way all the equilibrium values are ordered (which corresponds
to the width of the energy gap that determines the equilibration)
tells us the peculiarity of the Redfield scheme, namely, that the
reorganization energies are mixed according to the \textquotedbl{}initial\textquotedbl{},
i.e. optical energy gap (cf. Eq. \eqref{eq:mix_angle}). Hence, the
initial excitonic configuration is maintained throughout the whole
evolution.

This is even more strongly pronounced in the secular Redfield scheme
(Fig. \ref{fig:main evolutions}e and f) where the equilibration is
fully determined by the optical energy gap $\Delta\epsilon$ alone.
This shows us how severely the neglect of the interplay between the
populations and coherences can distort the picture of thermalization.
For instance, it would suggest that dimers with identical absorption
maxima positions but different line-widths would have identical equilibrium
population distributions. The full Redfield equations, on the contrary,
demonstrate that the optical energy gap, as the most intuitive input
parameter, along with the resonance coupling are not enough for the
description of thermal equilibrium in the case of different reorganization
energies. Moreover, even the F\"{o}rster energy gap can be a poor benchmark
for assessing the equilibrium populations (cf. Fig. \ref{fig:different Forster gaps}
bottom panel).

The HQME solutions (Fig. \ref{fig:main evolutions}c and d) bear some
resemblance to the results that follow from the Redfield equations.
However, an unexpected feature is the difference between equilibrium
population values for the case when $\lambda_{a}=\lambda_{b}$, which
again tells us about different energy gaps in each case. This points
to the action of the bath upon the system. And indeed we find that
the conventional excitonic basis is just an approximate eigenbasis
due to the non-vanishing coherences. Establishing the preferred basis
as given previously led us to conclude that a good measure of the
action of the bath is an effective resonance coupling $J_{eff}$,
which has been used to replace the original $J$ value with to reproduce
the effective energy gap defined by the thermal equilibrium in the
eigenbasis. It appears (cf. Fig. \ref{fig:polaron basis}) that the
connection between the original resonance coupling and the effective
one resembles the so-called small polaron transformation \citep{A.H.Romero1999}.
The details of such dynamic suppression of the resonance coupling
captured by the HQME is beyond the scope of this paper and are presented
elsewhere \citep{Gelzinis2011}.

Discussing the relaxation to the ground state requires a couple of
remarks about the excitonic mixing of lifetimes. From Eqs. \eqref{eq:excitonic rates}
or \eqref{eq:excitonic rates approx} we notice that the mixing of
lifetimes does not depend on which of the two states is the lower
one. In the special case of one state being extremely long-lived,
we can obtain the following picture: for a dimer degenerate in energies
both excitonic states would have lifetimes twice the lifetime of the
short-lived state in the site representation, whereas for different
site energies in the limit of negligible coupling $J$ we retrieve
the original values of the lifetimes. Now, the intermediate regime
(energy gap not too big and/or rather strong resonance coupling) tells
us that the shorter lifetime varies rather moderately, but the longer
one can acquire a huge range of values depending on the actual arrangement
of the system as demonstrated in Fig. \ref{fig:Mixing-of-lifetimes}.

The full picture of the excitation dissipation depends essentially
on the interplay between the relaxation to the ground state and the
thermalization driven by the system-bath interaction. The total process,
which is a two-exponential decay, is shown in Fig. \ref{fig:all 4 relaxations HQME}.
The exponents can be attributed to the thermalization and relaxation
to the ground state (cf. Eq. \eqref{eq:app C solution}). In the long-time
limit, both the lower and the higher state populations decay with
the same rate (the latter not shown here) because the equilibration
is much faster than the relaxation. Therefore, the systems relaxes
while being in a dynamic equilibrium. The reorganization energies
partly determine the rates of the initial step of evolution and the
amplitudes of the long-time decay. The long-time rates are identical
for all four combinations of $\lambda$'s (as seen from the parallel
slopes in the figure). These dependencies are qualitatively similar
to those of the evolutions without relaxation to the ground state.
Upon comparison of the numerical results with analytical expressions,
it can be seen that the faster relaxation rate largely determines
the rate of the whole process, and therefore the lifetime of the short-lived
excitonic state is the determining factor of the excitation relaxation
to the ground state. Moreover, as can be seen in Fig. \ref{fig:Solutions-of-HQME+R}
and as follows from Eqs. \eqref{eq:app C eigenvalues}, in the presence
of the system-bath interaction the decay rates are different depending
on which of the two states is the lower one (cf. the case of purely
excitonic mixing of lifetimes). The decay is faster if the short-lived
state is below the long-lived one, however, for the parameters used,
the difference is just of a few picoseconds.

The parameters of the heterodimer chosen for calculations are similar
to those of the Chl\textendash{}Car dimer, which is assumed to be
responsible for the additional quenching of the excitation under the
NPQ conditions \citep{S.Bode2009}. Indeed, the excitation lifetime
of the $S_{1}$ state of carotenoids is short (less than $10\,\textrm{ps}$)
and the optical transition from the ground state to the $S_{1}$ state
is forbidden, hence the excitation and reorganization energies in
this case are not directly observable. However, the reorganization
energy is a decisive parameter of the system-bath coupling, therefore,
we have to consider the possible influence of the reorganization energy
of the Car molecule on the excitation relaxation in such a heterodimer.
It might be expected, that the excitation energy of the Car molecule
in a specific Chl\textendash{}Car dimer should be slightly above the
excitation energy of the Chl molecule under unquenched conditions,
and tends to be lower under the NPQ ones \citep{berera-kennisPNAS06}.
This statement can be examined using our model considerations, which
demonstrate that it is only partially valid. The short lifetime component
is the dominant control parameter of the excitation lifetime in such
a dimer independently of the relative positioning of the molecular
excitations of the monomers. As shown in Fig. \ref{fig:Solutions-of-HQME+R}
the excitation decays with the characteristic lifetime of $8\,\textrm{ps}$
if the optically allowed excited state of the Chl molecule is higher
than the optically forbidden excited state of Car. It decays with
the lifetime of $13\,\textrm{ps}$ in the opposite case. Fast (of
the order of $100\,\textrm{fs}$) relaxation of the excitation from
the initially excited state of the Chl molecule to the excited state
of Car taking place in the former case determines the main difference
between both situations. Thus, the difference in the quenching efficiency
between these two cases are related to the difference in the rates
of excitation transfer back to the chlorophylls of the surrounding
antenna. Moreover, the dominant effect of the quenching ability of
such heterodimer is determined not by the excitonic effects (see Fig.
\ref{fig:Mixing-of-lifetimes}), but rather by the exciton-bath interaction.
As can be concluded from Eqs. \eqref{eq:app C eigenvalues}, the molecule
characterized by the shortest relaxation time forms the valve of the
excitation relaxation and defines the excitation lifetime within the
heterodimer.

The same model of heterodimer is also applicable to exciton\textendash{}CT
state coupling conditions \citep{Renger2004,T.Mancal2006,T.Mancal2008}.
Since the lifetimes of both states are long and comparable, the dynamics
within such a dimer corresponds to the limit of ${\cal K}=0$. The
quenching ability of this dimer is determined solely by the thermalization
within the excitonic states. However, Fig. \ref{fig:main evolutions}b
(or d) demonstrates that, under certain configuration of reorganization
energies, the CT state can be the higher one (red curve) even though
it should be the lower one to ensure the quenching ability in the
F\"{o}rster limit.

\section{Conclusions}

We have considered dynamics within a heterogeneous system of two interacting
molecules. The main sources of heterogeneity in this system are the
difference in excitation energies, asymmetry in reorganization energies
(spectral line-widths) and the different excited state lifetimes of
the chromophores. We point out that difference of reorganization energies
introduces an ambiguity in the concept of energy gap between the excited
states. This situation is particularly important when considering
differences between coherent and incoherent transfer, since upon excitonic
mixing under certain combinations of reorganization energies, the
two states can become energetically swapped with respect to the uncoupled
ones. Furthermore, employing the HQME technique revealed that the
conventional excitonic basis becomes renormalized under the action
of the bath.

The analytical study of excitonic mixing of the excited state lifetimes
and the estimates following from it revealed that the lifetime of
the long-living state can be indeed greatly reduced. However, we found
that the final picture of energy dissipation from the system crucially
depends on the system-bath interaction rather than just on the excitonic
coupling. Namely, the short-living state, irrespectively of its energetic
position, forms the valve of the process and largely determines the
rate of dissipation by setting its lower boundary. We conclude that
in the case of the Chl\textendash{}Car dimer the rate of excitation
relaxation is of the similar order as the inverse lifetime of Car
$S_{1}$ state even if the state is above the Chl excited state. This
has direct consequences in assessing the model of Chl\textendash{}Car
dimer as a quenching center of the NPQ process.

\section*{Acknowledgment}

This research was partly funded by the European Social Fund under
the Global Grant Measure. T.M. acknowledges the support from the Ministry
of Education, Youth and Sports of the Czech Republic through grants
KONTAKT 899 and MSM0021620835. V.B. thanks the Research Council of
Lithuania for the support of his three months stay at Charles University
in Prague.

\appendix

\section{Transformation of the Relaxation Superoperator}

The unitary transformation of a superoperator follows from the definitions
of the analogous transformation of an operator and the action of a
superoperator, accordingly:

\begin{equation}
\tilde{\rho}=U^{-1}\rho U\;\Longrightarrow\tilde{\rho}_{ab}=\sum_{ij}(U^{-1})_{ai}\rho_{ij}U_{jb};\label{eq:app B operator}
\end{equation}

\begin{equation}
\tilde{\rho}={\cal U}^{-1}\rho\;\Longrightarrow\tilde{\rho}_{ab}=\sum_{cd}({\cal U}^{-1})_{ab,cd}\rho_{cd}.\label{eq:app B superoperator}
\end{equation}
This way any superoperator ${\cal O}$ can by transformed as $\tilde{{\cal O}}={\cal U}^{-1}{\cal O}{\cal U}$.
Upon comparison of \eqref{eq:app B operator} and \eqref{eq:app B superoperator}
we can immediately deduce a rule for composing the unitary transformation
superoperator:

\begin{equation}
({\cal U}^{-1})_{ab,cd}=(U^{-1}\rho U)_{ab}\vert_{\rho_{ij}=\delta_{ic}\delta_{jd}}.\label{eq:app rule}
\end{equation}
For instance, in the case of a heterodimer, we use Eq. \eqref{eq:uni_transf}
to obtain the following superoperator:

\begin{alignat}{1}
{\cal U}^{-1} & =\left(\begin{array}{cccc}
c^{2} & sc & sc & s^{2}\\
-sc & c^{2} & -s^{2} & sc\\
-sc & -s^{2} & c^{2} & sc\\
s^{2} & -sc & -sc & c^{2}
\end{array}\right);\label{eq: super-transformation}
\end{alignat}
here we use the shorthand notations: $c=\cos\theta$, $s=\sin\theta$.
Therefore, bearing in mind that ${\cal K}_{ab,ab}={\cal K}_{ba,ba}=\frac{\kappa_{a}+\kappa_{b}}{2}$,
it is straightforward to show that the result of a product ${\cal U}^{-1}{\cal K\,}{\cal U}$
for the population relaxation elements yields Eqs. \eqref{eq:excitonic rates}.


\begin{thebibliography}{10}
\bibitem{A.V.Ruban2011}
A.~V. Ruban, M.~P. Johnson, C.~D.~P. Duffy, The photoprotective molecular
  switch in the photosystem ii antenna., Biochim. Biophys. Acta.

\bibitem{M.G.Mueller2010}
M.~M\"uller, P.~Lambrev, M.~Reus, E.~Wientjes, R.~Croce, A.~R. Holzwarth, Chem
  Phys Chem 11 (2010) 1289--1296.

\bibitem{P.Horton1996}
P.~Horton, A.~Ruban, R.~Walter, Annual Review of Plant Physiology and Plant
  Molecular Biology 47 (1996) 655--684.

\bibitem{holt-flemingScience05}
N.~E. Holt, D.~Zigmantas, L.~Valkunas, X.-P. Li, K.~K. Niyogi, G.~R. Fleming,
  Carotenoid cation formation and the regulation of photosynthetic light
  harvesting, Science 307 (2005) 433--436.

\bibitem{T.K.Ahn2008}
T.~K. Ahn, T.~J. Avenson, M.~Ballottari, Y.~C. Cheng, K.~K. Niyogi, R.~Bassi,
  G.~R. Fleming, Science 320 (2008) 794--797.

\bibitem{S.Bode2009}
S.~Bode, C.~C. Quentmeier, P.-N. Liao, N.~Hafi, T.~Barros, L.~Wilk, F.~Bittner,
  P.~J. Walla, Proc. Nat. Acad. Sci. USA 106 (2009) 12311--12316.

\bibitem{vanGrondelle-NATURE2007}
A.~V. Ruban, R.~Berera, C.~Ilioaia, I.~H.~M. {van Stokkum}, J.~T.~M. Kennis,
  A.~A. Pascal, H.~{van Amerongen}, B.~Robert, P.~Horton, R.~{van Grondelle},
  Nature 450 (2007) 575.

\bibitem{berera-kennisPNAS06}
R.~Berera, C.~Herrero, I.~H.~M. {van Stokkum}, M.~Vengris, G.~Kodis, R.~E.
  Palacios, H.~{van Amerogen}, R.~{van Grondelle}, D.~Gust, T.~A. Moore, A.~L.
  Moore, J.~T.~M. Kennis, A simple artificial light-harvesting dyad as a model
  of excess energy dissipation in oxygenic photosynthesis, Proc. Nat. Acad.
  Sci. USA 103 (2006) 5343--5348.

\bibitem{Liao2011}
P.-N. Liao, S.~Pillai, D.~Gust, T.~A. Moore, A.~L. Moore, P.~J. Walla,
  Two-photon study on the electronic interactions between the first excited
  singlet states in carotenoid-tetrapyrrole dyads, J. Phys. Chem. A 115 (2011)
  4082--4091.

\bibitem{Davydov-book}
A.~Davydov, A Theory of Molecular Excitions, Mc.Graw-Hill, New York, 1962.

\bibitem{RashbaExcitons}
E.~I. Rashba, M.~D. Sturge (Eds.), Excitons, Elsevier, 1987.

\bibitem{Silinsh1994}
E.~A. Silinsh, V.~Capek, Organic Molecular Crystals. Interaction, Localization
  and Transport Phenomena, AIP Press, New York, 1994.

\bibitem{valkunasbook}
H.~{van Amerogen}, L.~Valkunas, R.~{van Grondelle}, Photosynthetic {E}xcitons,
  World Scientific, Singapore, 2000.

\bibitem{May2004}
V.~May, O.~K\"uhn, Charge and energy transfer in molecular systems, Wiley-VCH,
  2004.

\bibitem{ChoFleming2005}
M.~Cho, G.~R. Fleming, The {I}ntegrated {P}hoton {E}cho and {S}olvation
  {D}ynamics. {II}. {P}eak {S}hifts and 2{D} {P}hoton {E}cho of a {C}oupled
  {C}hromophore, J. {C}hem. {P}hys. 123 (2005) 114506.

\bibitem{Kjellberg2006}
P.~Kjellberg, B.~Br\"uggemann, T.~Pullerits, Two-dimensional electronic
  spectroscopy of an excitonically coupled dimer, Phys. Rev. B 74~(2) (2006)
  024303.

\bibitem{Psliakov-Fleming-JCP2006}
A.~V. Pisliakov, T.~Man\v{c}al, G.~R. Fleming, Two-dimensional optical
  three-pulse photon echo spectroscopy. ii. signatures of coherent electronic
  motion and exciton population transfer in dimer two-dimensional spectra, J.
  Chem. Phys. 124 (2006) 234505.

\bibitem{Abramavicius2010}
D.~Abramavicius, V.~Butkus, J.~Bujokas, L.~Valkunas, Manipulation of
  two-dimensional spectra of excitonically coupled molecules by
  narrow-bandwidth laser pulses, Chemical Physics 372 (2010) 22--32.

\bibitem{G.S.Schlau-Cohen2011}
G.~S. Schlau-Cohen, A.~Ishizaki, G.~R. Fleming, Two-dimensional electronic
  spectroscopy and photosynthesis: Fundamentals and applications to
  photosynthetic light-harvesting, Chem. Phys. 386 (2011) 1--22.

\bibitem{Ishizaki2009}
A.~Ishizaki, G.~R. Fleming, On the adequacy of the redfield equation and
  related approaches to the study of quantum dynamics of electronic energy
  transfer, J. Chem. Phys. 130 (2009) 234110.

\bibitem{Renger2004}
T.~Renger, Phys. Rev. Lett. 93 (2004) 188101.

\bibitem{T.Mancal2006}
T.~Man\v{c}al, L.~Valkunas, G.~R. Fleming, Chem. Phys. Lett. 432 (2006)
  301--305.

\bibitem{Grondelle2006}
R.~van Grondelle, V.~I. Novoderezhkin, Energy transfer in photosynthesis:
  experimental insights and quantitative models, Phys. Chem. Chem. Phys. 8
  (2006) 793--807.

\bibitem{HeijsKnoester2007}
D.~J. Heijs, A.~G. Dijkstra, J.~Knoester, Ultrafast pump-probe spectroscopy of
  linear molecular aggregates: Effects of exciton coherence and thermal
  dephasing, Chem. Phys. 341 (2007) 230--239.

\bibitem{T.Mancal2008}
T.~Man\v{c}al, L.~Valkunas, E.~L. Read, G.~S. Engel, T.~R. Calhoun, G.~R.
  Fleming, Spectroscopy 22 (2008) 199--211.

\bibitem{Rui-XueXu2009}
R.-X. Xu, B.-L. Tian, J.~Xu, Q.~Shi, Y.~Yan, Hierarchical quantum master
  equation with semiclassical drude dissipation, J. Chem. Phys. 131 (2009)
  214111.

\bibitem{mukbook}
S.~Mukamel, Principles of {N}onlinear {O}ptical {S}pectroscopy, Oxford
  University Press, New York, 1995.

\bibitem{ishizaki:234111}
A.~Ishizaki, G.~R. Fleming, Unified treatment of quantum coherent and
  incoherent hopping dynamics in electronic energy transfer: Reduced hierarchy
  equation approach, The Journal of Chemical Physics 130~(23) (2009) 234111.

\bibitem{olsina-2010}
J.~Ol\v{s}ina, T.~Man\v{c}al, Electronic coherence dephasing in excitonic
  molecular complexes: Role of markov and secular approximations, J. Mol.
  Model. 16 (2010) 1765.

\bibitem{Gelzinis2011}
A.~Gelzinis, D.~Abramavicius, L.~Valkunas, Non-markovian effects in
  time-resolved fluorescence spectrum of molecular aggregates: tracing polaron
  formation, Phys. Rev. B (2011) submitted.

\bibitem{Schlosshauer2007}
M.~Schlosshauer, Decoherence and the Quantum-to-Classical Transition, The
  Frontiers Collection, Springer Verlag, Berlin, Germany, 2007.

\bibitem{A.H.Romero1999}
A.~H. Romero, D.~W. Brown, K.~Lindenberg, Phys. Rev. B 59 (1999) 13728--13740.

\end{thebibliography}
\end{document}